\documentclass[%
reprint,
 amsmath,amssymb,
 aps,
]{revtex4-1}

\usepackage{graphicx}% Include figure files
\usepackage{dcolumn}% Align table columns on decimal point
\usepackage{bm}% bold math
\usepackage{graphics}
\usepackage{amsmath}
\usepackage{amssymb}
\usepackage{mathrsfs}
\usepackage{amsfonts}
\usepackage{fancyhdr}
\usepackage[T1]{fontenc}
\usepackage{float}
\usepackage{verbatim}
\usepackage{pdfpages}
\usepackage{hyperref}
\usepackage{lmodern}
\usepackage{default}
\usepackage{url}
\usepackage{subfigure}
\usepackage{epstopdf}

\begin{document}
\bibliographystyle{unsrt}

\preprint{APS/123-QED}

%\title{Manuscript Title}% Force line breaks with \\

\title{Active Brownian agents with concentration-dependent chemotactic sensitivity}
%\\Pattern Formation and Dynamics}
%\subtitle{Pattern Formation and Dynamics}

\author{Marcel Meyer}
% \affiliation{Department of Physics, Newtonstr 15, 12489 Berlin}%Lines break automatically or can be forced with \\
\author{Lutz Schimansky-Geier}%
\affiliation{Department of Physics, Humboldt Universit\"at zu Berlin, Newtonstr 15, 12489 Berlin}%Lines break automatically or can be forced with \\

\author{Pawel Romanczuk}%
\affiliation{Physikalisch-Technische Bundesanstalt, Abbestr. 2-12, 10587 Berlin}%Lines break automatically or can be forced with \\

 % \affiliation{

%\author{Marcel Meyer\inst{1} \and Lutz Schimansky-Geier\inst{1} \and Pawel Romanczuk\inst{2}}% etc
%}                     % Do not remove
%\institute{Department of Physics, Newtonstr 15, 12489 Berlin \and Physikalisch-Technische Bundesanstalt, Abbestr. 2-12, 10587 Berlin}


%\affiliation{
%\affiliation{
%\affiliation{%


\date{\today}% It is always \today, today,

\begin{abstract}
We study a biologically motivated model of overdamped, autochemotactic Brownian agents with concentration-dependent chemotactic sensitivity. The agents in our model move stochastically and produce a chemical ligand at their current position. The ligand concentration obeys a reaction-diffusion equation and acts as a chemoattractant for the agents, which bias their motion towards higher concentrations of the dynamically altered chemical field. 
We explore the impact of concentration-dependent response to chemoattractant gradients on large-scale pattern formation, by deriving a coarse-grained macroscopic description of the individual based model, and compare the conditions for emergence of inhomogeneous solutions for different variants of the chemotactic sensitivity. We focus primarily on the so-called ``Receptor Law'' sensitivity, which models a nonlinear decrease of chemotactic sensitivity with increasing ligand concentration. Our results reveal qualitative differences between the  ``Receptor-Law'', the constant chemotactic response and the so-called ``Log-Law'', with respect to stability of the homogeneous solution, as well as the emergence of different patterns (labyrinthine structures, clusters and bubbles)  via spinodal-decomposition or nucleation. We discuss two limiting cases, where the model can be reduced to the dynamics of single species: (I) the agent density governed by an density-dependent effective diffusion coefficient and (II) the 
ligand-field with an effective bistable, time-dependent reaction rate. In the end, we turn to single clusters of agents, studying domain growth and determining mean characteristics of the stationary inhomogeneous state. Analytical results are confirmed and extended by large-scale GPU simulations of the individual based model.
%\begin{description}
%\end{description}
\end{abstract}

\pacs{Valid PACS appear here}% PACS, the Physics and Astronomy
\maketitle

%\tableofcontents

\section{Introduction}
Chemotaxis is defined as the directed motion of cells along chemical concentration gradients (see e.g. \cite{Murray2007}). It enables individual cells to bias their motion towards 
favorable environmental conditions and is therefore essential for the survival of a plethora of bacterial species. Furthermore, it is an important principle in the 
dynamics of various other biological systems, e.g.: guidance of leukocyte cells in the process of wound healing \cite{Lauffenburger1988,Murray2007}, cancer cell invasion 
\cite{Hatzikirou2010,Werbowetski2004,Quaranta2008}, and neuronal self-wiring \cite{Segev2000,Hentschel1999}. In cases where the corresponding chemicals are 
produced by the cells themselves as a reaction to environmental conditions, chemotaxis has been interpreted as effective cell-to-cell communication \cite{Benjacob1997}. Such autochemotactic response, 
which may be attractive or repulsive, plays an important role in the aggregation and internal dynamics of bacterial colonies (see e.g. \cite{Budreneberg1995,
Budreneberg1991,Benjacob2000,Benjacob1994}). Very recently it was shown, that also artificial systems of self-propelled colloids, may resemble the behaviour of living cell by exhibiting 
chemotactic drift and chemical signaling \cite{Hong2007,Theurkauff2012}. 

Since the pioneering work of Patlak, Keller and Segel \cite{Patlak1953,Kellersegel1970}, a huge progress has been made on developing and analysing various models of chemotaxis. 
Most modeling approaches are based on partial differential equations (PDEs) for the density of cells and concentration of the chemoattractants and/or repellents (see e.g. 
\cite{Tyson1999a,Tindall2008}). 
But also various individual based models (IBMs) were suggested, which reflect the discrete and stochastic nature of the modelled system. A major advantage of such models is the possibility 
to introduce individual features of cell behaviour directly into the mathematical model. The downside is the possible difficulties in deriving a coarse-grained description 
in terms of a small number of PDEs, which allows analytical predictions on stability and large-scale behaviour of the system. One should also note, that with respect to numerical integration, individual based 
models have some advantages over corresponding PDEs as they naturally account for the low-density limit and may be 
less susceptible to instabilities in situations with strong density inhomogeneities. In particular for IBMs it is possible to massively reduce simulation times by using Graphical Processing Units (GPUs), 
due to the intrinsically parallel 
hardware of GPUs. General purpose computing on GPUs
has been employed by a growing number of scientists during the last decade. For models of non-interacting particles, speed-ups by factor of 600 were reported \cite{Januszewski2010}. 
A review on GPU-computation in the field of statistical 
physics is given in \cite{Preis2011}. We designed an
optimized simulation setup on GPUs for the efficient numerical integration of our IBM even at high densities.

In this work, we will introduce and study a biologically motivated model of active Brownian agents \cite{Romanczuk2012} with a concentration-dependent chemotactic sensitivity, modelled by the so-called \emph{Receptor Law} (RL); The RL models a non-linearly decreasing chemotactic sensitivity with increasing ligand concentration -- a phenomenon observed in different bacteria species, as well as neuronal growth cones \cite{Tindall2008,Murray2007,Segev2000,Lapidus1976,Lauffenberger1984,Widman1997,Chiu2001}.
The introduced model is related to our previous work in individual based modelling of auto-chemotactic agents \cite{LSGS1994,Romanczuk2008}. Other microscopic models of (auto-) chemotaxis with constant chemotactic sensitivity have been studied in \cite{Grima2004,Grima2005,Senguptalowen2009,Taktikos2012,Chavanis2010}. Our microscopic model with a concentration-dependent chemotactic sensitivity, reduces in a limiting case to the constant-sensitivity model originally studied by Schweitzer and Schimansky-Geier in 1994 \cite{LSGS1994,Schweitzer2003}. 
Furthermore, in 
a different limiting case, it may be related to the so-called ``Log-law'' for chemotactic sensing.  The nonlinear chemotactic drift term discussed here, was previously
 used in a model of self-propelled particles in combination with additional local velocity-alignment interaction \cite{Romanczuk2008}. We consider here an overdamped model of chemotactic Brownian agents, in order to focus specifically on the impact of the RL on macroscopic pattern formation. Hereby, we show that in our model the stability properties of the homogeneous solutions as well as the emergent structures are qualitatively different from the constant chemotactic response model introduced in  \cite{LSGS1994}.  

After introducing our model (Section \ref{modelintro}), we will proceed with the analysis of the corresponding coarse-grained equation. Here, we will perform a linear stability analysis of the homogeneous solution 
for different variants of chemotactic sensitivity and show, in particular, that for a chemotactic drift according to the RL, the homogeneous state  is linearly stable at high and low particle densities. At intermediate densities the homogeneous solution becomes unstable and the system exhibits a wide range of transient spatio-temporal patterns, which for $t \to \infty$ converge towards a stable inhomogeneous state (Section \ref{patternformation}). 
Two limiting cases of the model will be discussed: (1) Fast relaxation of the particle density and (2) fast relaxation of the chemical concentration field 
(Section \ref{limitcases}). At the end, we will focus on the inhomogeneous state, discussing domain growth and mean stationary characteristics of single clusters (Section
\ref{singleclusters}). 

\section{Model setup: Concentration dependent chemotaxis of active Brownian agents}
\label{modelintro}
\subsection{Microscopic model equations} 
We consider an ensemble of active autochemotactic Brownian agents indexed with $i=1,2,\dots, N$, as a model for bacteria that interact via 
chemotaxis. 
Different species of bacteria, as e.g. \emph{Escherichia coli} (E. coli), produce chemoattractants (or chemorepellents) to interact  and ``communicate'' e.g. favorable or hostile environments 
\cite{Romanczuk2008,Murray2007}. 
Motivated by this observation, agents in our model produce a chemical ligand $c({\bf r},t)$ at their time-dependent position ${\bf r}_i(t)$ with a constant production rate $q_c$.
The concentration field of the chemical $c({\bf r},t)$ obeys a reaction-diffusion equation and acts as a chemoattractant for the agents: They move towards higher concentrations 
of the dynamically altered chemical field. The movement of agents in our model is characterized by the interplay of two forces in the Langevin-equations governing 
individual trajectories: (I) the chemotactic force ${{\bf F} }_{\text{chem}}$ determining the response of agents to ligand field and (II) a stochastic force ${\bf F}_{\text{stoch}}$ modeling 
the random movement of agents. We consider a 2-dimensional system with $N$ point-like agents with mass $m=1$ in the overdamped limit. This brings us to the following 
microscopic model equations for $i=1\dots N$ Brownian agents:
\begin{subequations}
\label{ibmmodel}
\begin{align} 
\label{modelposition}
%\frac{\partial{\bf r}_i}{\partial t}&=\frac{\kappa}{\gamma\left(1+\beta c({\bf r}_i,t)\right)^2} \frac{\partial c({\bf r},t)}{\partial r}\bigg|_{{\bf r}_i} + \sqrt{2D}\xi_i(t)
\frac{d{\bf r}_i}{d t}&=\frac{1}{\gamma}{\bf F}_{\text{chem}}({\bf r}_i,t) + \sqrt{2D}{\boldsymbol{\xi}}_i(t),\\
\label{modelfield}
\frac{\partial c({\bf r},t)}{\partial t}&=q_c\sum_{i=1}^N{\delta({\bf r}-{\bf r}_i(t))}-d_cc({\bf r},t)+ D_c \Delta c({\bf r},t).
\end{align}
\end{subequations}
The first term on the right side of Eq. \ref{modelposition} describes the overdamped, chemotactic drift with a constant friction coefficient $\gamma$, and a force-term 
${\bf F}_{\text{chem}}$ which will be specified in the next 
paragraph (Eq. \ref{rlforce}). The second term models the random movement of agents with noise intensity $D$ and Gaussian white noise vector ${\boldsymbol{\xi}}_i(t)$. 
Equation \ref{modelfield} describes the evolution of the chemical: We assume that the chemical is produced with a constant production rate $q_c$ at the position $r_i(t)$ of every agent, decomposes with a decay rate $d_c$ and diffuses with a diffusion coefficient $D_c$. 

Biological agents, as e.g. many forms of bacteria, are able to sense local concentrations of a ligand by measuring the relative occupation, $\sigma_i$, of its membrane receptors, which for independent receptors can
 be expressed as
\begin{align}
 \sigma_i=\frac{N_o}{N_o+N_f}=\frac{\tau_o}{\tau_o+\tau_f}.
\end{align}
Here $N_o$ describes the number of occupied receptors and $N_f$ the number of free receptors, which are assumed to be proportional to the mean occupation time $\tau_o$ and 
the mean free time $\tau_f$ of the receptors, respectively \cite{Benjacob1997,Segev2000,Murray2007}. Following \cite{Benjacob1997,Segev2000,Romanczuk2008}, we assume that the unbinding rate of the receptor $k_{o\to f}=\tau_o^{-1}$ is independent of the ligand concentration $c$. Thus there exists a constant characteristic mean occupation time:  $\tau_o=const.$. On the other hand, we assume the binding rate of the receptors $k_{f\to o}=\tau^{-1}_f$ to be directly proportional to the ligand concentration. This yields the free time $\tau_f$ to be inversely proportional to the space- and time-dependent concentration of the field, $\tau_f=\frac{a_o}{c({\bf r},t)}$ (where $a_o$ 
is a constant).

Eukaryotic cells are sufficiently large to sense differences in ligand concentration along the cell body \cite{Ueda2007,Amselem2012a,Amselem2012b}. Their chemotactic response 
can be assumed directly proportional to the spatial gradient of $\sigma_i$. Bacterial cells, on the other hand, are in general too small to detect spatial gradients directly, 
and have to rely on temporal sensing of differences in receptor occupation. However, assuming a finite stochastic displacement of biological agents per unit time 
$|\delta {{\bf r_i}}/\delta t|$, combined with the ability to sense temporal changes in $\sigma_i$, leads to the effective measurement of spatial gradients of the relative occupation of receptors on length scales larger than the persistence length of cell motion\cite{Erban2004,Romanczuk2008}. Thus on sufficiently large length scales chemotactic bacteria can effectively be described as Brownian agents with a net drift along gradients due to an effective chemotactic force. 

Based on above considerations, we assume the chemotactic force to be proportional to the spatial gradient of the relative occupation of membrane receptors and the corresponding time constants, respectively 
\begin{align}
\label{propchem}
 {\bf F}_{\text{chem}}\propto \nabla \frac{N_o}{(N_o+N_f(c({\bf r}_i,t))}\propto\nabla \frac{\tau_0}{(\tau_o+\tau_f(c({\bf r}_i,t))}.
\end{align}
This allows us to write the effective chemotactic force acting on each agent $i$ as a function of the concentration of the chemical 
\begin{align}
\label{receptorlaw}
 {\bf F}_{\text{chem}}({\bf r}_i,t)=\frac{\chi_o K}{\left(K+c({\bf r}_i,t)\right)^2}\nabla c({\bf r}_i,t)
\end{align}
where we calculated the gradient in Eq. \ref{propchem} and introduced $\chi_o$ as a proportionality factor and $K=a_o/\tau_o$ as a constant determined by the characteristic mean time of receptor occupation.
The resulting prefactor of the gradient on the right side of Eq. \ref{receptorlaw} is known as the receptor law \cite{Murray2007,Segev2000,Tindall2008,Lapidus1976,Lauffenberger1984,Widman1997,Chiu2001}. 
Already decades ago it was pointed out that best agreement between experimental results and models has been obtained for a concentration-dependent chemotactic sensitivity of the form of the 
receptor law \cite{Lapidus1976}. Throughout the past decades, it has been used for modelling of bacteria in continuous macroscopic models by e.g. 
\cite{Lapidus1976,Lauffenberger1984,Widman1997,Chiu2001}, but most of the microscopic approaches have so far only considered constant chemotactic drift. 
By extending the right side of Eq. \ref{receptorlaw} with $\frac{1/K^2}{1/K^2}$, defining the chemotactic drift coefficient $\kappa=\frac{\chi_o}{K}$, and the chemotactic saturation parameter $\beta=\frac{1}{K}$, we arrive at an alternative formulation of the receptor law drift
\begin{align}
\label{rlforce}
 {\bf F}_{\text{chem}}({\bf r}_i,t)=\underbrace{\frac{\kappa}{\left(1+\beta c({\bf r}_i,t)\right)^2}}_{\chi^{RL}(c)}\nabla c({\bf r}_i,t)
\end{align}
with $\kappa,\beta\ge 0$. For $\beta>0$ the effect of the RL concentration-dependent chemotactic sensitivity $\chi^{RL}(c)$ can qualitatively be sketched as follows: For low $c({\bf r},t)$ ($\ll \beta^{-1}$) the denominator of Eq. \ref{rlforce} is approximately $1$ and agents are maximally sensitive to gradients of the chemoattractant. 
However, if the concentration of the chemical increases, the denominator of the drift-term grows and the chemotactic force decreases nonlinearly. 
In the biological picture, $c \gg \beta^{-1}$ corresponds to situations where almost all membrane receptors are occupied and agents become insensitive to local chemoattractant gradients.
The rescaled version of the RL in Eq. \ref{rlforce} was used in \cite{Romanczuk2008} and explicitly distinguishes between (I) chemotactic drift (determined by $\kappa$) and (II) chemotactic saturation (determined by $\beta$).
We use it here for an intuitive comparison between the model of (i) constant chemotactic sensitivity ($\beta=0$) studied in \cite{LSGS1994} and (ii) concentration-dependent chemotactic 
sensitivity studied in this work. 

In the limiting case $\beta=0$ (or rescaled according to Eq. \ref{receptorlaw}: $c\ll K$), the receptor law sensitivity $\chi^{RL}(c)$ reduces to constant chemotactic sensitivity 
\begin{align}\label{constforce}
 \chi^{o}=\kappa.
\end{align}

The receptor law is derived by assuming independent cellular receptor dynamics, but it has been shown for E. coli bacteria, that adaptation dynamics of the intra-cellular chemotactic signalling system violate these assumptions and that the chemotactic response in E. coli strains is better described by the so-called Log-Law sensitivity \cite{Kalinin2009}:
\begin{align}
 \label{loglaw}
 \chi^{LL}(c)=\frac{\kappa}{c}.
\end{align}
Following \cite{Lauffenberger1984,Murray2007} it can be obtained as a special case of the RL by considering $c\approx K$ (or rescaled: $\beta^{-1}\approx c$ and $\kappa^{-1}\approx c$).
In Section \ref{patternformation} we will compare the collective behaviour that results from the three chemotactic sensitivities $\chi^0$, $\chi^{LL}$ and $\chi^{RL}$, but further on we will focus on RL sensitivity.

The autochemotactic coupling between the agents implies a positive feedback between the number of agents and the concentration of the field $c({\bf r},t)$: The higher the number of agents in a certain region, the more chemicals are produced and therefore even more agents are attracted. In the special case of $\beta=0$, positive feedback finally leads to a ``blow up'' of the system with all agents collapsing into a single delta-peaked distribution \cite{LSGS1994}. 
In the next Sections, we will show how a finite saturation parameter $\beta>0$ effects the collective behaviour of the Brownian agents.

\subsection{Macroscopic description and global behaviour}
In order to analyze the system at large length-scales, it is advantageous to consider the coarse-grained density of particles $\rho({\bf r},t)$, rather than the microscopic, stochastic description of single agents.
We assume the factorization of the $N$-particle probability distribution function (PDF) into a product of $N$ one-particle PDFs (mean-field assumption). This allows us to derive directly
a Smoluchovski -- equation for the one-particle PDF $P({\bf r},t)$, corresponding to our overdamped Langevin equations 
(Eq. \ref{modelposition}). Further on, we will describe the system in terms 
of the one-particle density function $\rho({\bf r},t)=NP({\bf r},t)$, where $N$ is the total number of agents. 
Corresponding coarse-graining has to be performed with respect to the production term of the reaction-diffusion equation governing $c({\bf r},t)$ (Eq. \ref{modelfield}); Here, we arrive at 
the macroscopic perspective via (see also \cite{LSGS1994,Romanczuk2008}):
\begin{align}
 q_c\sum_{i=1}^N{\delta({\bf r}-{\bf r}_i(t))}\rightarrow q_cNP({\bf r},t)\rightarrow q_c\rho({\bf r},t).
\end{align}
This brings us to the following macroscopic model description for the RL-sensitivity: 
\begin{subequations}
\label{continoummodel}
\begin{align}
\label{makrodichte}
\frac{\partial \rho({\bf r},t)}{\partial t}&=\nabla \Big(-\frac{\kappa\rho({\bf r},t) }{\gamma(1+\beta c({\bf r},t))^2} \nabla c({\bf r},t) + D \nabla \rho({\bf r},t) \Big)\\
%\frac{\partial \rho({\bf r},t)}{\partial t}&=\nabla \Big(-\frac{{\bf F}_\text{chem}}{\gamma}\rho({\bf r},t) \nabla c({\bf r},t) + D \nabla \rho({\bf r},t) \Big)\\
\frac{\partial c({\bf r},t)}{\partial t}&=q_c\rho({\bf r},t)-d_c c({\bf r},t)+D_c \Delta c({\bf r},t). 
\label{makrofeld}
\end{align}
\end{subequations}
This set of continuous partial differential equations can be seen as a special realization of the well known Patlak-Keller-Segel model \cite{Patlak1953,Kellersegel1970} with a receptor-law chemotactic sensitivity.  

Throughout this work, we will use the following integral operator to represent the global average of an arbitrary function $g({\bf r},t)$:
\begin{align}
\langle g \rangle (t) & = \frac{1}{A} \int_A g({\bf r},t) d{\bf r}\ ,
\end{align}
wit $A$ being the total area of the system.

For no flux, or periodic boundary conditions (as used in the simulations), the diffusion of the chemical does not change its total amount in the system (as $N=\langle \rho \rangle A = const.$). The time-dependent global mean chemoattractant concentration $\langle c \rangle(t)$ therefore obeys
\begin{align}
\frac{d\langle c \rangle (t)}{dt}&=q_c \langle \rho \rangle -d_c \langle c \rangle(t) .
\end{align}
The solution of the above ODE reads
\begin{align}
\label{totalconcentration}
 \langle c \rangle(t)&=\frac{q_c}{d_c}\langle \rho \rangle \left(1-e^{-d_ct} \right).
\end{align}
The total amount of chemoattractant increases continuously and for $t>5\tau$ with $\tau=\frac{1}{d_c}$, it reaches more than $99\%$ of its final amount and thus can be considered constant
\begin{align}
\label{globalfield}
 %C_{tot}\overset{t > 5\tau}{=}\frac{q_c}{d_c}N.
 \lim_{t\to \infty} \langle c \rangle (t)=\frac{q_c}{d_c}\langle \rho \rangle.
\end{align}
We can therefore distinguish between two temporal regimes with respect to the global amount of chemoattractant in the system: At short times, for $0<t<5\tau$, the system is characterized by a global growth of $\langle c \rangle (t)$, whereas at long times and for $t>5\tau$ only local redistribution of $c({\bf r},t)$ takes place, with the global amount of chemoattractant being well approximated by its constant stationary limit given in Eq. \ref{globalfield}. If not stated otherwise, we will focus on the case of stationary global field ($t\gg 5\tau$) with local redistribution. 
%For $t\rightarrow\infty$ the system evolves into a stationary inhomogeneous state. 

In the following Sections, we will study the stability of the homogeneous solution of Eq. \ref{makrodichte} and \ref{makrofeld} and the behaviour of possible inhomogeneous states. Analytic results will be compared to GPU-simulations of the microscopic model equations \ref{modelposition} and \ref{modelfield}.

\subsection{GPU Simulation Setup}
The Langevin-equations governing the single trajectories of agents were numerically integrated on GPU's
based on a combination of optimized algorithms introduced in \cite{Green2008} and \cite{Januszewski2010} and the
reaction-diffusion equation was calculated using the GPU finite differences algorithm described in \cite{Micikevicius2009}. 
For systems with large ensembles of particles and large simulation grids this simulation setup leads to significant 
speed-ups. No explicit benchmarking was done as we concentrated on the characteristics of our model, but an exemplary comparison gives an idea of the powerful computational capabilities: 
The GPU (NVIDIA Tesla M2050) needs 60 seconds for $10^4$ discrete integration s of $4\cdot10^5$ agents on a $512\times512$ simulation grid, while the CPU (DELL intel i7-3770) needs 2758 seconds. 
This corresponds to a speed-up by a factor of 46. For smaller systems of 1000 agents on a $256\times256$ grid, we observed only a speed-up by a factor of 2. All simulations described in this work where conducted in 2D systems on regular squared simulation grids (grid length set to 1) with periodic boundary conditions.

\section{Pattern formation}
\label{patternformation}
\subsection{Linear stability analysis}
In this Section, we perform a linear stability analysis of the macroscopic equations \ref{makrodichte} and \ref{makrofeld}, and compare the results for the RL-sensitivity with the corresponding results for LL (Eq. \ref{loglaw}) and constant sensitivity (Eq. \ref{constforce}). Hereby, we show that RL sensitivity leads to two density regimes (high and low $\langle \rho \rangle$), where the homogeneous solution remains stable, and to an unstable regime at intermediate $\langle \rho \rangle$, where different patterns emerge.

A simple solution of the macroscopic model equations \ref{makrodichte} and \ref{makrofeld} is the stationary and spatially homogeneous solution 
%\begin{subequations}
 \begin{align}
 \label{homogensolution}
 \overline \rho =\langle \rho \rangle, \qquad
 \overline c =\lim_{t\to\infty}\langle c \rangle(t)\ . 
\end{align}
%\end{subequations}
In general, $\langle \rho \rangle=N/A$ is an important parameter of the system, and only in the particular case of a homogeneous system it represents also a solution of the macroscopic 
equations. We note that Eq. \ref{globalfield} yields a simple relation between the homogeneous density and concentration
\begin{align}
\label{homogensimple}
 \overline c={q_c\overline \rho}/{d_c}
\end{align}
which will be frequently used.

Considering a spatially infinite system, we allow small fluctuations around the stationary homogeneous solution (Eq. \ref{homogensolution}) 
\begin{subequations}
 \label{perturbations} 
\begin{align}
\rho({\bf r},t)&=\overline \rho + \delta \rho,\quad \text{with} \quad \delta \rho=ilon \sum_kf_k(t)e^{-i{\bf k}{\bf r}}\\
c({\bf r},t)&=\overline c + \delta c,\quad \text{with} \quad \delta c=ilon \sum_kg_k(t)e^{-i{\bf k}{\bf r}} 
\end{align}
\end{subequations}
with $f_k(t)$ and $g_k(t)$ being the Fourier amplitudes of the wave vector ${\bf k}$ of the perturbations  and $ilon\ll1$. 
Using the ansatz $e^{\lambda t}$ and Eq. \ref{globalfield}, we derive the following dispersion relation:
%\begin{align}
%\label{dispersionrelation}
%  &\lambda_{+/-}({\bf k})=-\frac{\left((D_c+D){\bf k}^2+d_c\right)}{2} \quad \pm \qquad \\ 
% \scriptsize & \sqrt{\frac{\left((D_c+D){\bf k}^2+d_c\right)^2}{4} - \left(DD_c{\bf k}^2 + Dd_c - \frac{\kappa\overline{\rho}q_c}{\gamma\left(1+\beta\frac{q_c}{d_c}\overline{\rho}\right)^2}\right){\bf k}^2} . \nonumber
%\end{align}
\begin{widetext}
\begin{align}
\label{dispersionrelation}
  \lambda_{+/-}({\bf k})=-\frac{\left((D_c+D){\bf k}^2+d_c\right)}{2} \pm  
  \sqrt{\frac{\left((D_c+D){\bf k}^2+d_c\right)^2}{4} - \left(DD_c{\bf k}^2 + Dd_c - \frac{\kappa\overline{\rho}q_c}{\gamma\left(1+\beta\frac{q_c}{d_c}\overline{\rho}\right)^2}\right){\bf k}^2} .
\end{align}
\end{widetext}
%%%%%%%%%%%%%%%%%%%%%%%%%%% FIGURE %%%%%%%%%%%%%%%%%%%%%%%%%%%%%%%%%%%
\begin{figure*}
\begin{center}
\includegraphics[width=0.8\textwidth]{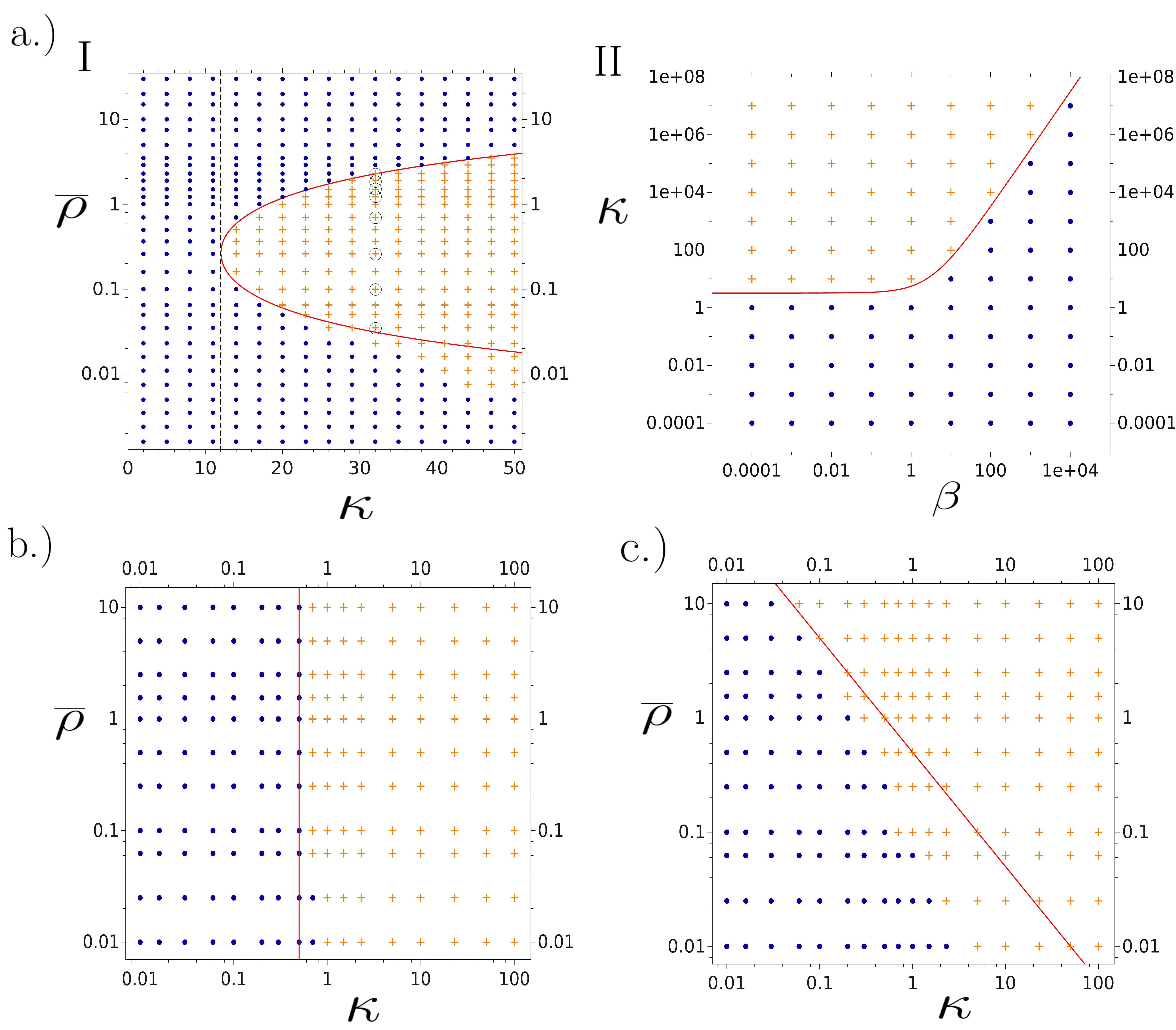}
\caption{ (Color online) Comparison of the results of linear stability analysis (red lines) and simulations (blue circles: no pattern formation; orange crosses: pattern formation) for different chemotactic sensitivities $\chi(c)$.
(a) RL sensitivity $\chi^{RL}(c)=\frac{\kappa}{(1+\beta c)^2}$. 
(a.I) Stability for different values of $\kappa$ and $\overline{\rho}$. Red line: $\rho{\pm}$ (see Eq. \ref{bistablekrit}). Dashed line: $\kappa=4D\gamma\beta$ 
(see Eq. \ref{stabcond}). Simulation parameters: $\beta=1.5$, $D=2$, $\gamma=d_c=1$, $q_c=2.5$, $D_c=5$, $\Delta t=1\cdot10^ {-3}$;  
Black circles for $\kappa=32$ mark simulations for different densities with snapshots shown in Fig. \ref{transientstrukturen} . 
(a.II) Stability with RL sensitivity for different 
chemotactic parameters $\kappa,\beta>0$. Red line: $\kappa_{crit}$ according to Eq. \ref{kcrit}. Simulation parameters: $D=\gamma=q_c=D_c=1,d_c=0.1$; 
(b) Stability for log-law chemotactic sensitivity $\chi^{LL}(c)=\frac{\kappa}{c}$. Red line shows the results of Eq. \ref{loglawbed}. 
(c) Stability for constant chemotactic sensitivity $\chi^o=\kappa$. Red line  shows the results of  Eq. \ref{constbed}. Simulation parameters for (b) and (c): $D=0.5,d_c=q_c=2,\gamma=1,D_c=0.1$.}
\label{stabsimulations}
\end{center}
\end{figure*}
%%%%%%%%%%%%%%%%%%%%%%%%%%%%%%%%%%%%%%%%%%%%%%%%%%%%%%%%%%%%%%%%%%%%%%%%%%%%%%%%%%%%%%
For inhomogeneous fluctuations (finite ${\bf k}$-vectors) the homogeneous solution is 
only stable as long as
\begin{align}
\label{stabbedingung}
 \left(DD_c{\bf k}^2+Dd_c-\frac{\kappa\overline\rho\ q_c}{\gamma\left(1+\beta\frac{q_c}{d_c}\overline \rho \right)^2}\right) \ge 0.
\end{align}
Considering the limit of perturbations with long wavelength ${\bf k}\rightarrow 0$ and solving for $\overline \rho$ in order to obtain a
 stability condition for the density of particles yields
\begin{align}
\label{parabel}
 \overline \rho^2 + \overline \rho \frac{d_c}{q_c\beta}\left(2-\frac{\kappa}{D\gamma\beta}\right)+\frac{d_c^2}{\beta^2 q_c^2} \ge 0 \ .
\end{align}
The two solutions, for which the left term of Eq. \ref{parabel} is equal to zero, are
\begin{align}
\label{bistablekrit}
 \rho_{\pm} 
 %& = \frac{d_c}{q_c\beta}\left[\left(\frac{\kappa}{2D\gamma\beta}-1\right)\pm\sqrt{\left(\frac{\kappa}{2D\gamma\beta}-1\right)^2-1}\right] \nonumber \\  
            & =\frac{\kappa d_c}{2D\gamma\beta^2 q_c}\left[\left(1-\frac{2D\gamma\beta}{\kappa}\right)\pm\sqrt{1-\frac{4D\gamma\beta}{\kappa}}\right].
\end{align}
These are the critical densities at which the stability of the homogeneous solution changes. 
%For $\beta,q_c,d_c>0$ 
The condition
\begin{align}
\label{stabcond}
 \kappa>4D\gamma\beta
\end{align}
(or $\chi_o>4D\gamma$, if we rescale according to Eq. \ref{receptorlaw}) defines the parameter space in which the radicand in Eq. \ref{bistablekrit} is positive and we find two critical densities $\rho_{\pm}$ with real, positive values. 
In this case, we obtain the following stability conditions for the homogeneous solution $\overline \rho$
\begin{subequations}
\begin{align}
 \overline \rho &\le  \rho_- : \text{stable} \\
 \rho_-<\overline \rho &<   \rho_+ : \text{unstable} \\
\overline \rho &\ge  \rho_+ : \text{stable}. 
\end{align}
\end{subequations}
The concentration-dependent RL results in two regimes at high and low particle density, respectively, where the homogeneous solution is stable. In between, there exists an intermediate density regime, where we expect growth of inhomogeneities and emergence of patterns within the system. 
For a fixed density of particles, we solved Eq. \ref{stabbedingung} for $\kappa$ in order to explicitly express the stability with respect to the chemotactic drift coefficient $\kappa$ and the chemotactic saturation $\beta$. The homogeneous solution is stable against inhomogeneous fluctuations as long as 
\begin{align}
\label{kcrit}
 \kappa\le\kappa_{crit}=\frac{D\gamma d_c}{\overline\rho q_c}\left(1+\beta\frac{q_c}{d_c}\overline\rho\right)^2.
\end{align}
Thus, for $\kappa>\kappa_{crit}$ it is unstable and we expect to see pattern formation. In the special case of $D=0$ (and positive other model parameters), the system is unstable for all $\kappa>0$.
The homogeneous solution can only be stable for finite diffusion of the particle density.
 
In the special case of $\beta^{-1}\approx c$, the RL reduces to the ``Log-law'' (LL) sensitivity  
(compare e.g. \cite{Lauffenberger1984,Murray2007}). The linear stability analysis of Eq. \ref{makrodichte} and Eq. \ref{makrofeld} analogous to the above described, but with $\chi^{LL}(c)=\frac{\kappa}{c}$, leads to the following stability condition
\begin{align}
\label{loglawbed}
\kappa\le D\gamma.
\end{align}
While the RL induces two density regimes in which the homogeneous solution is stable, a LL sensitivity implies that linear stability of the homogeneous state is independent of the density of particles. 
For constant chemotactic sensitivity, as the second special case of the RL with $\beta=0$, we note that the homogeneous solution is stable as long as 
\begin{align}
\label{constbed}
\overline{\rho}\le\frac{D\gamma d_c}{q_c\kappa} 
\end{align}
and therefore depends on the density of particles but exhibits only one stable density regime at lower densities \cite{LSGS1994}.

We compared the above results of the linear stability analysis for the different cases of chemotactic sensitivity (RL, LL, constant) to results of microscopic simulations (see Fig. \ref{stabsimulations}). 
For each type of response, we initialized a set of systems with homogeneous particle densities varying over several orders of magnitude. Starting from the lowest 
densities of $\overline{\rho}=0.0016$ (26 agents) up to $\overline{\rho}=30$ ($4.9\cdot10^5$ agents), we evaluated the pattern formation at different values of 
the chemotactic drift coefficient $\kappa$. Please note, the extremely low number of agents in the lowest density case, makes the coarse-graining approach questionable 
and therefore the numerical results of the individual based simulations are likely to deviate from the analytic predictions resulting from the macroscopic field equations.  

As shown in Figure \ref{stabsimulations}-(a.I) numerical calculations for RL sensitivity confirm that $\rho_{\pm}$ (Eq. \ref{bistablekrit}) distinguishes two regimes of the particle density in which the homogeneous solution is stable and an intermediate regime in which we observe growing inhomogeneities of $\rho({\bf r},t)$ and $c({\bf r},t)$.
The deviation of the numerical results from the analytical description for lower densities in Figure \ref{stabsimulations}-(a.I) can be understood by keeping in mind that we integrated the microscopic equations of motion for the single
trajectories of the agents: At low particle densities this can easily lead to fluctuations of the density which are beyond the linear regime.
At a fixed density of $\overline{\rho}=0.031$ we tested the analytical expression for $\kappa_{crit}$ described in Eq.\ref{kcrit}. Figure \ref{stabsimulations}-(a.II) shows the very good agreement between analytical and numerical results concerning the interdependence of the chemotactic drift coefficient $\kappa$  and the chemotactic saturation parameter $\beta$. Here, we emphasize that, in contrast to the model with self-propelled particles \cite{Romanczuk2008}, the matching between theory and simulation results is obtained without any fit parameter. 

Figures \ref{stabsimulations}b and \ref{stabsimulations}c illustrate stability for LL sensitivity and constant chemotactic drift. Numerical simulations confirm that for $\chi(c)=\frac{\kappa}{c}$ stability is independent of $\overline{\rho}$ and determined by Eq. \ref{loglawbed}. For constant chemotactic drift, we validate the stability condition given by Eq. \ref{constbed}, but note that for lower densities of particles, we see deviations from the analytical predictions: Without  chemotactic saturation, nonlinear feedback between $\rho({\bf r},t)$ and $c({\bf r},t)$ during growth of inhomogenities is increased in comparison to the RL sensitivity, where the chemotactic saturation 
reduces sensitivity with growing $c({\bf r},t)$. Thus, higher-order positional correlation become non-negligible for the microscopic dynamics at low densities, and the coarse-grained approximation becomes invalid. This results in large deviations of the simulation results from the predictions of the linear stability analysis.    
%%%%%%%%%%%%%%%%%%%%%%%%%%%%%%%% FIGURE %%%%%%%%%%%%%%%%%%%%%%%%%%%%%%%%%%%%%%%%%%%%%%%%%
\begin{figure}
\begin{center}
\includegraphics[width=0.96\columnwidth]{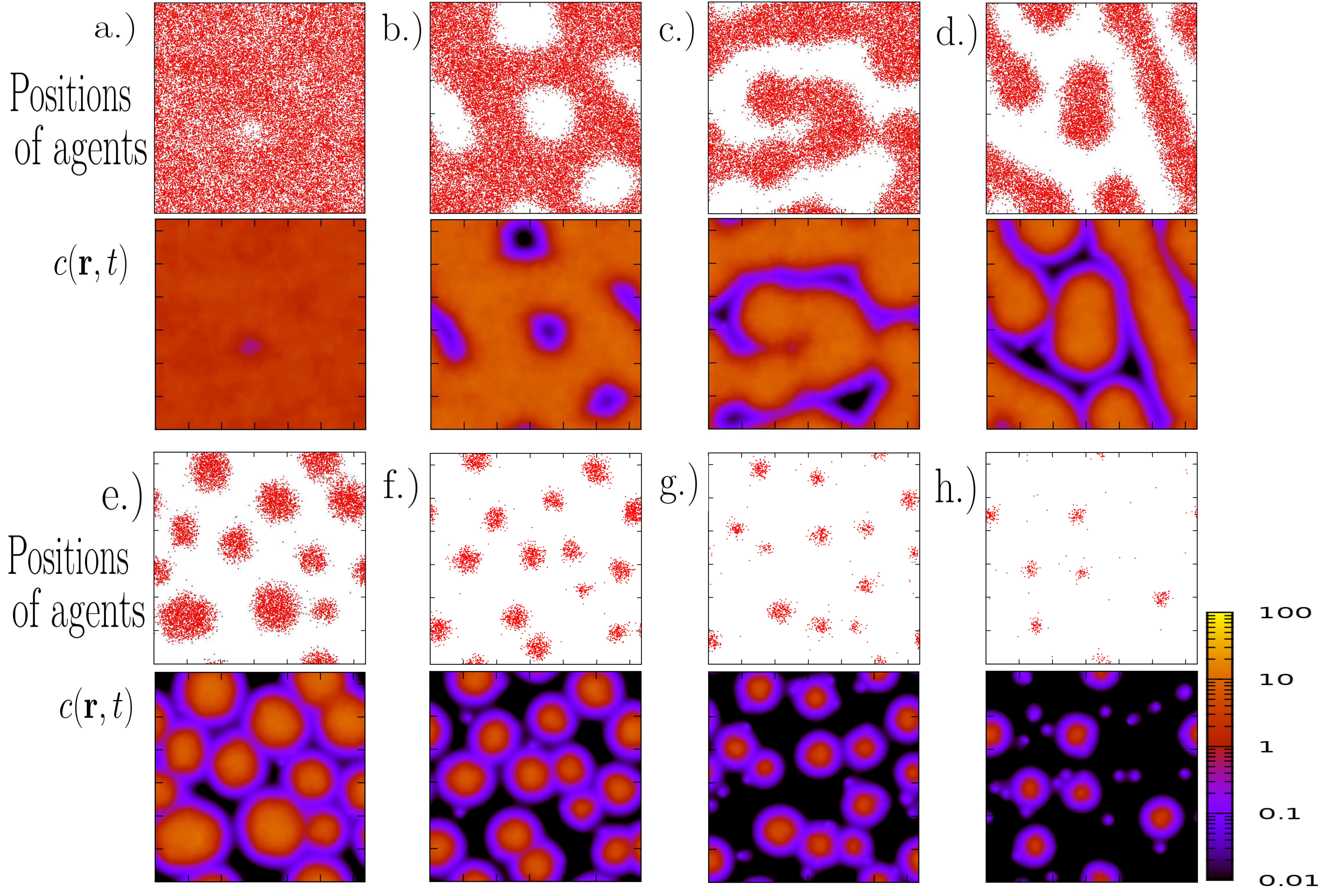}
\caption{ (Color online) Transient patterns emerging with RL chemotactic sensitivity. All snapshots taken at the same simulation time from the simulations shown in Figure \ref{stabsimulations}-(a.I) 
(on a $128\times128$ simulation grid) for different 
mean densities of particles at a fixed value $\kappa=32$. 
(a) $\rho$=2.3; (b) $\rho$=1.9; (c) $\rho$=1.5; (d) $\rho$=1.22; (e) $\rho$=0.7; (f) $\rho$=0.26; (g) $\rho$=0.1; (h) $\rho$=0.035.} 
\label{transientstrukturen}
\end{center}
\end{figure}
%%%%%%%%%%%%%%%%%%%%%%%%%%%%%%%%%%%%%%%%%%%%%%%%%%%%%%%%%%%%%%%%%%%%%%%%%%%%%%%%%%%%%%%%%%%%
\subsection{Transient Patterns} 
In the unstable regime, we observe a variety of transient patterns. Figure \ref{transientstrukturen} illustrates a selection of typical transients for different densities of particles with RL sensitivity. 
All snapshots are taken from the simulations in Figure \ref{stabsimulations}-(a.I) for different homogeneous initial densities $\overline{\rho}$ at the same simulation time and at a fixed value of $\kappa=32$. Close to the critical line $\rho_+$ in the unstable regime we observe the formation of \emph{bubbles} (see Figure \ref{transientstrukturen}a and \ref{transientstrukturen}b). 
Decreasing the density of particles the bubbles get larger and formation of \emph{labyrinthine} structures can be observed (see Figure \ref{transientstrukturen}c). 
With further decrease of the particle density, the labyrinthine structures disappear and we see a transition to large, rather asymmetric 
\emph{clusters} (see Figure \ref{transientstrukturen}d). At low densities the agents 
form radially symmetric \emph{clusters} (see Figure \ref{transientstrukturen}e-h). The formation of clusters is experimentally well known from the aggregation of bacterial colonies as e.g. shown in \cite{Budreneberg1991}.
For very low values of $\kappa$, only slightly above $\kappa_{crit}$, we can observe initially the emergence of large regions of slightly higher density of particles instead of distinct clusters, which is indicative of a long-wavelength instability. 
Furthermore, for parameter values close to $\kappa_{crit}$ (Eq. \ref{kcrit}) single particles can dynamically move between clusters, and the structure of clusters can fluctuate strongly. 
Qualitatively this resembles the behaviour observed in the dynamical clustering of artificial colloids with chemical signaling studied in \cite{Theurkauff2012}.
We also see the analogous dynamics at high densities where insulated particles diffuse into bubbles. For parameters well above the 
critical line $\kappa_{crit}$, individual agents are stronger bound and the clusters (bubbles) show only small fluctuations around a symmetric disc-like shape.

We usually observed typical characteristics of nucleation processes at high densities (close to $\rho_+$) and at low densities (close to $\rho_-$): Many small inhomogeneities grow and decay before overcritical stable clusters (bubbles) emerge. In contrast, labyrinthine structures at intermediate densities start to grow immediately and are distributed over the entire simulation space - strongly reminding us of spinodal decomposition patterns. We therefore note that the overdamped, autochemotactic Brownian agents with RL sensitivity exhibit a non-equilibrium phase-transition, that qualitatively resembles equilibrium liquid -- vapor phase -- transitions. 

Similar structures have very recently been studied in systems of self-propelled Brownian agents with density dependent mobility (see \cite{Tailleur2008,Potapov2005,Fily2012,Stenhammar2013,Cates2013}) and experimental results in \cite{Buttinoni2013} show that active colloids moving via diffusiophoresis also exhibit similar phase transitions at high densities. 

For LL sensitivity and constant drift (Fig. \ref{stabsimulations}b \& Fig. \ref{stabsimulations}c), we observe neither spinodal decomposition patterns nor bubble formation via nucleation. Clusters are formed, which are more compact than those of agents with RL sensitivity and show less fluctuations, due to the absence of nonlinear saturation effects.  
\subsection{Long-Term behaviour}
%%%%%%%%%%%%%%%%%%%%%%%%%% FIGURE %%%%%%%%%%%%%%%%%%%%%%%%%%%%%%%%%%%%%
\begin{figure}
\begin{center} 
\includegraphics[width=0.96\columnwidth]{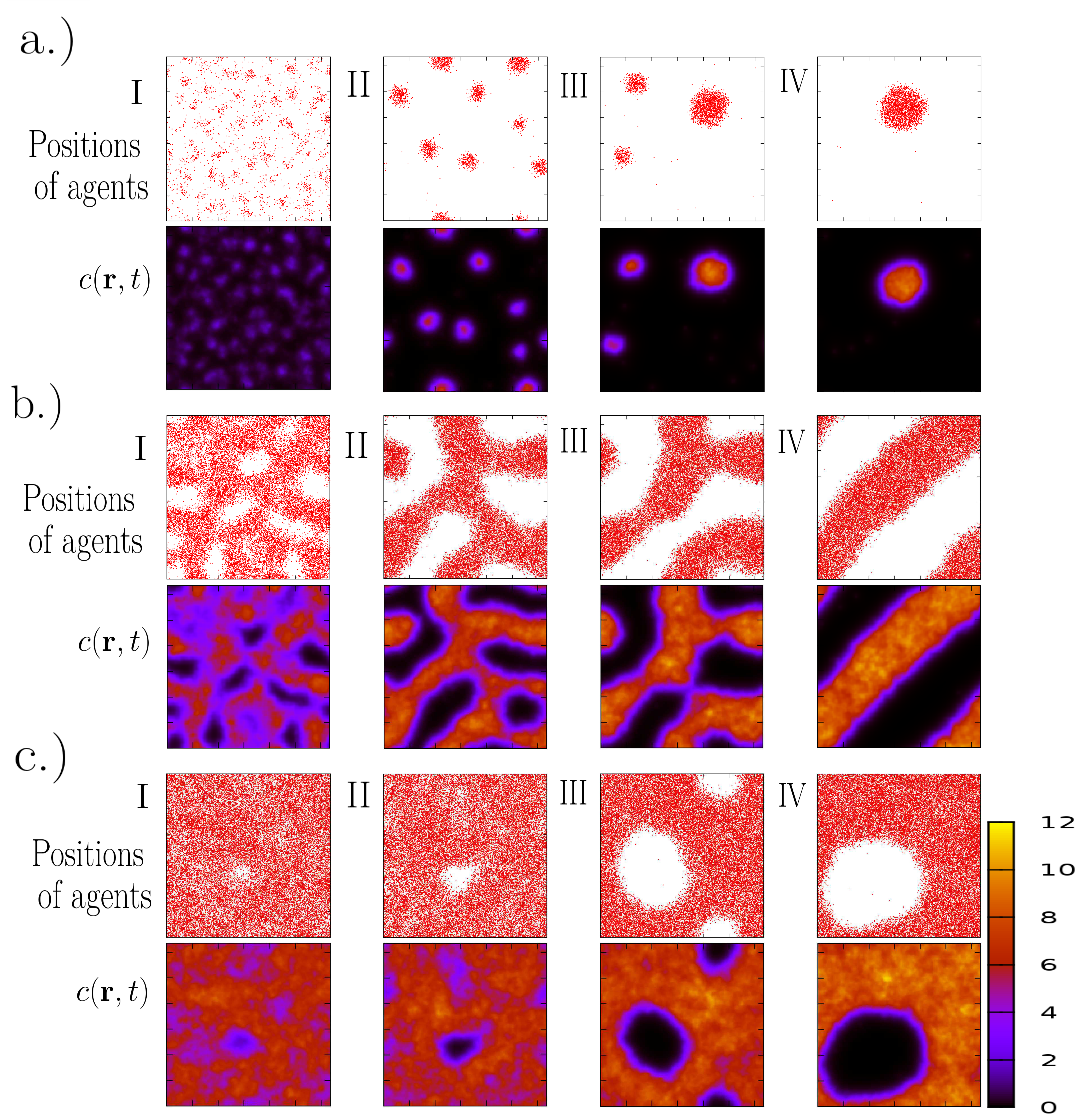}
\caption{ (Color online) Spinodal decomposition and nucleation of autochemotactic Brownian agents with RL sensitivity. Long-term behaviour at different mean densities of particles. 
(a) $\overline{\rho}=0.16$ at (I) $t=20$,   (II) $t=3000$, (III) $t=70000$, (IV) $t=3\cdot10^5$;
(b) $\overline{\rho}=1.5$ at (I)  $t=200$,  (II) $t=1000$, (III) $t=3000 $, (IV) $t=70000$;
(c) $\overline{\rho}=2.3$ at (I)  $t=1000$, (II) $t=1500$, (III) $t=3000 $, (IV) $t=70000$.
All in the same parameter regime as shown in Fig. \ref{stabsimulations}-(a.I) and Fig. \ref{transientstrukturen}. Numerical integration step: $\Delta t=5\cdot10^ {-3}$. Please note periodic boundary conditions in (b).}
 \label{longtimeplot}
\end{center}
\end{figure}
%%%%%%%%%%%%%%%%%%%%%%%%%%%%%%%%%%%%%%%%%%%%%%%%%%%%%%%%%%%%%%%
We will now focus on a system with RL sensitivity and describe its typical long-term behaviour at different particles densities. 
In the limit $t\rightarrow\infty$ we observe the decomposition of the diverse transient patterns shown in Figure \ref{transientstrukturen}. 
The system evolves to a stationary inhomogeneous state, characterized by one cohesive domain of high density of particles ($\rho>\rho_+$), a transition region and one domain of 
low density ($\rho<\rho_-$) of 
particles. 
Figure \ref{longtimeplot} shows the long-term behaviour of the model for three different initial homogeneous densities: (a) Low density rather close to the critical line $\rho_-$,
(b) intermediate densities and (c) high densities close to the critical line $\rho_+$.
Figure \ref{longtimeplot}-a illustrates the growth and merging of \emph{clusters} that eventually lead to a single surviving cluster at low mean density $\overline{\rho}=0.16$. Merging of clusters can be interpreted as Ostwald-Ripening \cite{LSGS1994,Romanczuk2008,Schweitzer1988} - it is the typical behaviour for systems with low mean density in the unstable parameter regime. 
At intermediate densities (Fig. \ref{longtimeplot}b) we see spinodal decomposition into two coexisting phases: The labyrinthine structures that evolve at $\overline{\rho}=1.5$ 
decompose and finally form one cohesive region of high density with a straight transition region. Figure \ref{longtimeplot}c shows a system of high mean 
density $\overline{\rho}=2.3>2.28=\rho_+$. 
Even though linear stability analysis predicts a stable homogeneous solution for $\overline{\rho}>\rho_+$, in this case, for a mean density only slightly above $\rho_+$, we see pattern 
formation via nucleation. At high densities we observe the growth and merging of \emph{bubbles} (instead of clusters at low densities) that eventually lead to one single bubble embedded in 
a region of high density of particles.
The concentration field $c({\bf r},t)$ in Figure \ref{longtimeplot} illustrates that many smaller 
inhomogeneities are formed in the region of high densities (especially at early times), but only a few become stable bubbles. 
Later in this paper we will come back 
to the systems shown in Fig. \ref{longtimeplot}).

\section{Discussion of limiting cases}
\label{limitcases}
In this Section we will assume that the concentration of the field $c({\bf r},t)$ and the 
particle density $\rho({\bf r},t)$ relax to its stationary state at significantly different time scales (similar approach in \cite{LSGS1994} for the model with $\beta=0$). 
This allows us to go beyond the results of the linear stability analysis. 
%%%%%%%%%%%%%%%%%%%%%%%%%%%% FIGURE %%%%%%%%%%%%%%%%%%%%%%%%%%%%%%%%%%%%%%%%%
\begin{figure*}
\begin{center}
\includegraphics[width=0.85\textwidth]{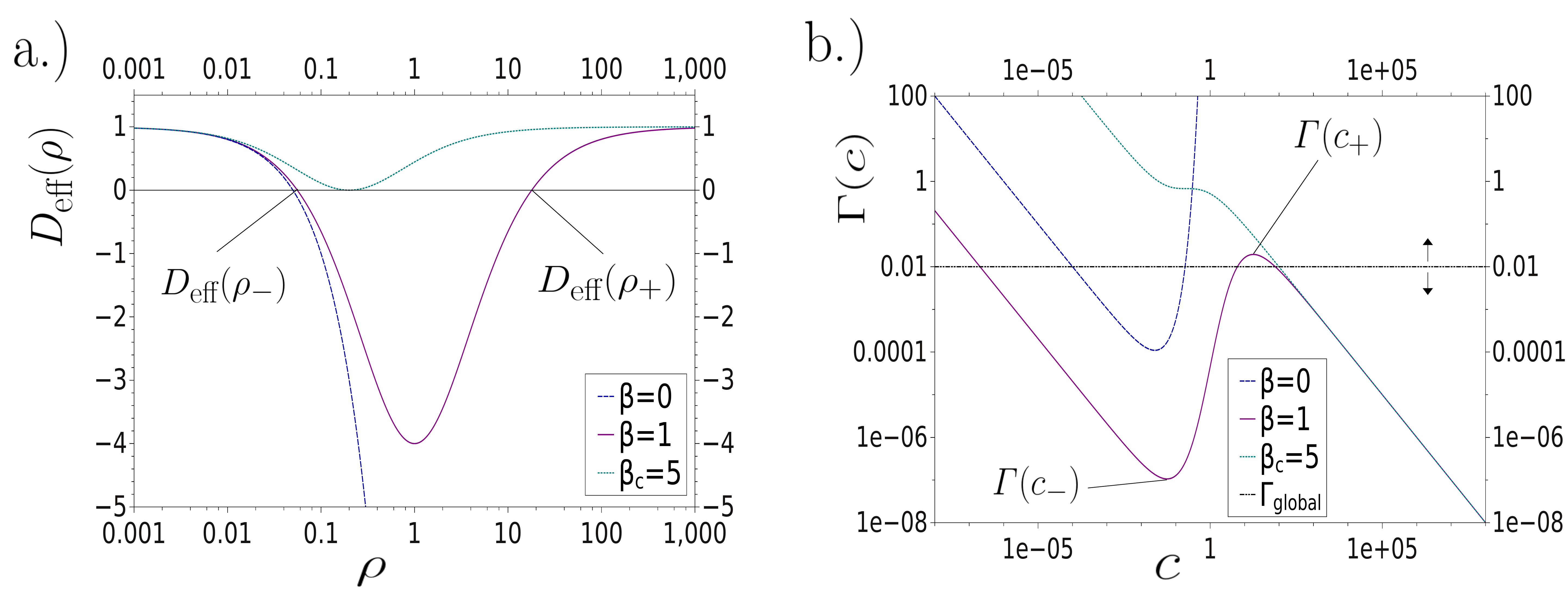}
\caption{ (Color online) Typical form of $D_{\text{eff}}(\rho)$ and $\Gamma(c)$ obtained in the limiting cases 
$\frac{\partial c}{\partial t}\approx0$ and $\frac{\partial \rho}{\partial t}\approx0$ for different regimes of chemotactic saturation $\beta$. (a) effective Diffusion coefficient 
$D_{\text{eff}}$ and (b) local fitness 
$\Gamma(c)$  compared to one exemplary value of the global fitness $\Gamma_{\text{global}}(c)$. Both (a) and (b) 
are plotted for the same set of parameters ($\frac{q_c}{d_c}=1, \kappa=20,D=\gamma=1$) and different values of the 
chemotactic saturation $\beta$
(the local fitness for the case $\beta=0$ is obtained from \cite{LSGS1994}). For $0<\beta<\beta_c=\frac{\kappa}{4D\gamma}$ (see Eq.\ref{stabcond}) the effective diffusion coefficient 
$D_{\text{eff}}$ enters the regime where it changes its sign and corresponding to that, two extrema appear in the local fitness. The densities $\rho_{\pm}$ and concentrations $c_{\pm}$ 
at which $D_{\text{eff}}(\rho)$ changes its sign and we find the extrema of $\Gamma(c)$ are connected by $c_{\pm}=\frac{q_c}{d_c}\rho_{\pm}$ (see Eq. \ref{cplusrhopluseq}) and are in accordance with the critical densities 
obtained from the linear stability analysis. }
\label{deffgrowth}
\end{center}
\end{figure*}
%%%%%%%%%%%%%%%%%%%%%%%%%%%%%%%%%%%%%%%%%%%%%%%%%%%%%%%%%%%%%%%%%%%%%

\subsection {Fast relaxation of the field}
\label{fastfield}
In the limiting case of a fast relaxation of the field into its stationary state, 
we assume that $c({\bf r},t)$ directly follows the particle density so that the time evolution of $\rho({\bf r},t)$ governs the coupled system. 
Assuming a quasi-stationary state of the field ($\frac{\partial c}{\partial t}\approx0$), allows us to rewrite the macroscopic density-Equation (Eq. \ref{makrodichte}) as an ordinary diffusion equation (as done in \cite{LSGS1994} for $\beta=0$):
\begin{align}
\label{diffequ}
 \frac{\partial \rho}{\partial t}&=\frac{\partial}{\partial {\bf r}}\left(D_{\text{eff}}({\bf r},t)\frac{\partial \rho}{\partial {\bf r}}\right)
\end{align}
with the \emph{effective diffusion coefficient}
\begin{align}
D_{\text{eff}}({\bf r},t)&=D-\frac{\kappa\rho({\bf r},t)}{\gamma(1+\beta c({\bf r}))^2}\frac{\partial c}{\partial \rho}.
 \end{align}
If we additionally consider the case of small diffusion of the field ($D_c\approx0$), fast relaxation of $c$ to its stationary state 
justifies the approximation, that at every position ${\bf r}$ the field $c$ is directly proportional to the particle density
\begin{align}
\label{local}
 c({\bf r},t)=\frac{q_c}{d_c}\rho({\bf r},t)\ .
\end{align}
Using this local relation allows us to simplify the effective diffusion coefficient and rewrite it as a function of the particle density 
\begin{align}
\label{effdiff}
 D_{\text{eff}}({\bf r},t)=D-\frac{\kappa \frac{q_c}{d_c}\rho({\bf r},t)}{\gamma\left(1+\beta  \frac{q_c}{d_c}\rho({\bf r},t)\right)^2}
\end{align}
or (using Eq. \ref{local}) as a function of the chemoattractant concentration, respectively.
Figure \ref{deffgrowth}a shows a plot of $D_{\text{eff}}(\rho)$ for different values of the saturation coefficient $\beta$. For $\beta=0$, the effective diffusion coefficient monotonically decreases 
with growing density of particles. However, for $0<\beta<\beta_c=\frac{\kappa}{4D\gamma}$ (see Eq. \ref{stabcond}) it is interesting to note, that $D_{\text{eff}}$ changes its sign at two 
densities $\rho_{\pm}$: For small $\rho<\rho_-$ it is positive, for intermediate densities $\rho_-<\rho<\rho_+$ it is negative and for high densities $\rho_+<\rho$ it is positive again. 
A positive effective diffusion coefficient corresponds to the spreading of the particle density while $D_{\text{eff}}<0$ leads to the agglomeration of particles. 
Setting $D_{\text{eff}}=0$ shows that the two densities, where the effective diffusion coefficient changes its sign, correspond to the densities $\rho_{\pm}$ obtained from the linear 
stability analysis of the homogeneous solution (see Eq. {\ref{bistablekrit}). The density regime in which the homogeneous solution is unstable corresponds to the density
regime where the effective diffusion coefficient is negative. Very recently it was shown in \cite{Stenhammar2013,Cates2013} that ABPs with density dependent motility can exhibit 
phase-transition characteristics that resemble those observed here.
 
Figure \ref{deffmischung}a shows the numerical calculation of the spatio-temporal evolution of the binary effective diffusion coefficient of a system initialized in the homogeneous state with parameters in the clustering regime. Ring-like 
regions of a negative effective diffusion coefficient can be observed in the outer regions of the clusters. 
Only in these local transition regions $\rho_-<\rho({\bf r},t)<\rho+$, we find particle densities (field concentrations) corresponding to  negative values of $D_{\text{eff}}$. Different to that, the effective diffusion coefficient for $\beta=0$ is strictly negative throughout entire clusters (see: \cite{LSGS1994}).  

%%%%%%%%%%%%%%%%%%%%%%%%%%%%%%% FIGURE %%%%%%%%%%%%%%%%%%%%%%%%%%%%%%%%%%%%%%%%%%%%%%%%%%%
\begin{figure}
\begin{center}
\includegraphics[width=0.9\columnwidth]{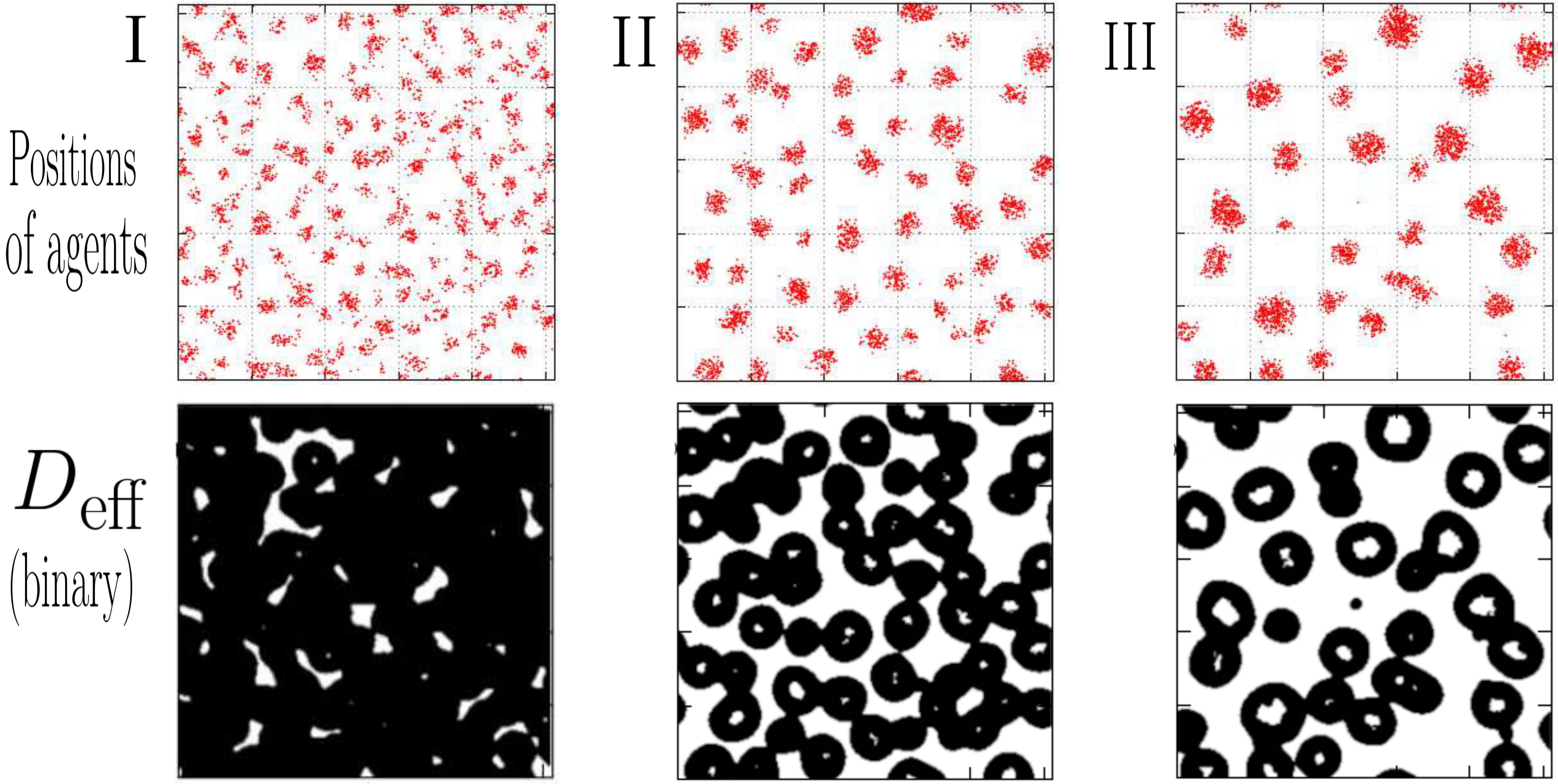}
\caption{ (Color online) Numerical calculation of the spatio-temporal evolution of the binary (black: negative, white: positive) effective diffusion coefficient $D_{\text{eff}}$ (based on Eq. 
\ref{effdiff} with $c=\frac{q_c}{d_c}\rho$ and $c({\bf r},t)$ from simulations). 
Time-s: (I) t=100, (II) t=500 and (III) t=4990 (simulation 
parameters: $\kappa=80,\beta=4,D=1.5,d_c=0.2,\gamma=q_c=D_c=1$).}
 \label{deffmischung}
\end{center}
\end{figure}
%%%%%%%%%%%%%%%%%%%%%%%%%%%%%%%%%%%%%%%%%%%%%%%%%%%%%%%%%%%%%%%%%%%%%%%%%%%%%%%%%%%%%%%%%%%%

\subsection {Fast relaxation of the particle density}
\label{fastdensity}
In the limiting case of a fast relaxation of the particle density, the time evolution of $c({\bf r},t)$ governs the coupled system. 
Following \cite{LSGS1994} we assume a quasi-stationary state of the particle density, $\rho_{\text{stat}}$ with $\frac{\partial \rho_{\text{stat}}}{\partial t}\approx0$. With 
respect to Eq. \ref{makrodichte} this yields
\begin{align}
  D\frac{\partial \rho_{\text{stat}}}{\partial {\bf r}}-\frac{\kappa\rho_{\text{stat}}}{\gamma(1+\beta c)^2}\frac{\partial c}{\partial {\bf r}}&=0
\end{align} 
for no flux, or periodic boundary conditions. The corresponding normalized quasi-stationary solution reads 
\begin{align}
 \label{normsol}
\rho_{\text{stat}}&=\overline{\rho}\frac{\exp\left(-\alpha(\beta+\beta^2c)^{-1}\right)}{\langle \exp\left(-\alpha(\beta+\beta^2c)^{-1}\right) \rangle} 
\end{align}
with $\alpha=\frac{\kappa}{D\gamma}$.

By plugging Eq. \ref{normsol} into Eq. \ref{makrofeld}, we obtain
 \begin{align}
  \label{growthfield}
\frac{\partial c({\bf r},t)}{\partial t}&=\underbrace{d_cc({\bf r},t)\left(\frac{\Gamma(c({\bf r},t))}{\Gamma_{\text{global}}(c({\bf r},t))\rangle}-1\right)}_{f(c({\bf r},t))} + D_c\Delta c({\bf r},t) \nonumber \\
&=f(c({\bf r},t)) + D_c\Delta c({\bf r},t).
\end{align}
Here we defined a reaction rate $f(c({\bf r},t))$ that is mainly determined by two terms: The numerator of the first term in brackets of Eq. \ref{growthfield} will be 
called \emph{local fitness} 
\begin{align}
\label {localgrowthrate}
  \Gamma(c({\bf r},t))=\frac{1}{c({\bf r},t)}\exp\left(-\frac{\alpha}{\beta+\beta^2c({\bf r},t)} \right),
 \end{align}
and the denominator we name \emph{global fitness}
\begin{align}
\label{globalgrowthrate}
 \Gamma_{\text{global}}(c({\bf r},t)) & =\frac{1}{\overline c}\left\langle \exp\left(-\frac{\alpha}{\beta+\beta^2c({\bf r},t)} \right) \right\rangle.
\end{align}
Hereby, we replaced $\overline \rho$ from Eq. \ref{normsol} using Eq. \ref{homogensimple}.

The names of the terms above are motivated by a structural analogy of Eq. \ref{growthfield} (for $D_c\approx0$) to the selection equations of the Eigen-Fisher type (see e.g. \cite{LSGS1994,Feistel2011}):
A single spot of high concentration (cluster of particles),  emerging in a system with low mean density can be interpreted as a species $j$, characterized by a fixed concentration $c_j$ instead of spatially extended concentration profile $c({\bf r},t)$, with a local fitness given by Eq. \ref{localgrowthrate}. 
From this perspective, time evolution of spots of high concentration represents a selection process. We will return to this interpretation further on, and proceed now with the analysis of the behaviour of Eqs. \ref{localgrowthrate}, \ref{globalgrowthrate} and \ref{growthfield}.

While $\Gamma(c({\bf r},t))$ can be different at every position ${\bf r}$ and every time step $t$ (corresponding to the local concentration $c({\bf r},t)$), the global fitness $\Gamma_{\text{global}}(c)$ is in general a time-dependent, global quantity, which depends on the spatial distribution of $c({\bf r},t)$ in the system.
As the global fitness depends on the spatial integral over the area $A$ of the system, it is reasonable to assume that it only changes slowly compared to a local changes in $c({\bf r},t)$. 
For a homogeneous distributions $\overline{c}$ (see Eq. \ref{homogensolution}), we note that 
\begin{align}
\label{globalhomogen}
\Gamma_{\text{global}}(\overline{c})
=\frac{1}{\overline{c}}\exp\left(-\frac{\alpha}{\beta+\beta^2\overline{c}}\right)=\Gamma(\overline{c}).
\end{align}

Considering small perturbations $\delta c({\bf r},t)$ around the homogeneous concentration $\overline{c}$, we linearize Eq. \ref{growthfield} around $\overline{c}$. With $\delta c\sim e^{\lambda t + i{\bf k}{\bf r}}$, we obtain
\begin{align}
\label{homogenstab}
 \lambda=\frac{d_c\overline{c}}{\Gamma(\overline{c})}\left(\frac{\partial \Gamma(c)}{\partial c}\bigg|_{c=\overline{c}} - \frac{\partial \Gamma_{\text{global}}(c)}{\partial c}\bigg|_{c=\overline{c}} \right) - {\bf k}^2D_c.
\end{align}
If we assume long wavelength perturbations, we can
neglect changes in the global fitness
\begin{align}
 \Gamma_{\text{global}}(\overline{c}+\delta c)  &\approx \Gamma_{\text{global}}(\overline{c}).
\end{align}
Therefore, stability around the homogeneous state depends only on the sign of the derivative of the local fitness.
Figure \ref{deffgrowth}b shows that for $0<\beta<\beta_{c}=\frac{\kappa}{4D\gamma}$ local fitness has one minimum at $c_-$ followed by one maximum at $c_+$. This implies the 
following stability conditions: 
\begin{subequations}
\begin{align}
\label{fitnessbedingung}
 \overline{c} &< c_- : \text{stable} \\
c_-<\overline{c} &< c_+ : \text{unstable} \\
\overline{c} &> c_- : \text{stable}. 
\end{align}
\end{subequations}
For small perturbations around $\overline{c}>c_+$ or $\overline{c}<c_-$, homogeneous solution is stable and global fitness remains stationary with $\Gamma_{\text{global}}(\overline{c})=\Gamma(\overline{c})$. But in the unstable regime for $c_-<\overline{c}<c_+$, fluctuations around the homogeneous distribution will grow, and according to the temporal evolution of the pattern, the global fitness
will change in time. Calculating the extrema of the local fitness, we get 
\begin{align}
\label{cplusrhopluseq}
  c_{\pm}=\frac{q_c}{d_c}\rho_{\pm}
\end{align}
where $\rho_{\pm}$ are the densities obtained from the linear stability analysis (see Eq. \ref{bistablekrit}), which correspond 
also to the values where the effective diffusion coefficient changes its sign. 
The correspondence of $\rho_{\pm}$ obtained from the effective diffusion coefficient and $c_{\pm}$ obtained from the
local fitness is illustrated in Fig. \ref{deffgrowth}. Figure \ref{deffgrowth}b shows a plot of the local fitness $\Gamma(c)$ for different values of the 
saturation coefficient $\beta$ compared to one exemplary value of the time-dependent global fitness $\Gamma_{\text{global}}(c)$. Please note that for the limit $\beta=0$, we used the corresponding expression from \cite{LSGS1994}.

Without chemotactic saturation ($\beta=0$) we see one minimum of $\Gamma(c)$ and only one density at which the effective diffusion coefficient changes its sign.
For $0<\beta<\beta_c$ on the other hand, there are two extrema $\Gamma(c_{\pm})$ and the concentrations $c_{\pm}$ correspond to the densities $\rho_{\pm}$ (via Eq. \ref{cplusrhopluseq}) where we find 
the two zeros of the effective diffusion 
coefficient 
$D_\text{eff}(\rho_{\pm})$.
With increasing chemotactic saturation $\beta$ the maximum of the local fitness $\Gamma(c_+)$ shifts towards lower concentrations of $c({\bf r},t)$. This corresponds to a shift of the critical value $\rho_+$ where the effective diffusion coefficient changes its sign. If on the other hand one increases $\kappa$, the maximum of $\Gamma(c)$, shifts towards higher values of $c({\bf r},t)$. For $\beta>\beta_c$ the local fitness monotonically decreases with increasing concentration $c$ and corresponding there is no change of sign of the effective diffusion coefficient (see Fig.\ref{deffgrowth}a).

Neglecting the diffusion of the field ($D_c\approx0$) in Eq. \ref{growthfield} allows us to get a closer look at the reaction rate $f(c)$ defined in Eq. \ref{growthfield}
\begin{align}
\label{reducedfield}
\frac{\partial c({\bf r},t)}{\partial t}=\frac{d_cc({\bf r},t)}{\Gamma_{\text{global}}(c({\bf r},t))}\Big(\Gamma(c({\bf r},t))- \Gamma_{\text{global}}(c({\bf r},t))\Big).
\end{align}
We can directly see that $f(c)>0$, if $\Gamma(c)>\Gamma_{\text{global}}(c) $ and $f(c)<0$ for 
$\Gamma(c)<\Gamma_{\text{global}}(c) $ (for $d_c>0$). The reaction rate is zero if $\Gamma(c)= \Gamma_{\text{global}}(c) $. For a locally vanishing concentration $c({\bf r},t)=0$ it approaches a finite value $f(c)=\frac{d_c}{\Gamma_{\text{global}}}\exp(-\alpha/\beta)$, which depends on the global fitness at the corresponding time. 
%%%%%%%%%%%%%%%%%%%%%%%%% FIGURE %%%%%%%%%%%%%%%%%%%%%%%%%%%%%%%%%%%%%%%%%%%%%%%%%%%%%%%%%%%%%%%%
\begin{figure}
\includegraphics[width=0.98\columnwidth]{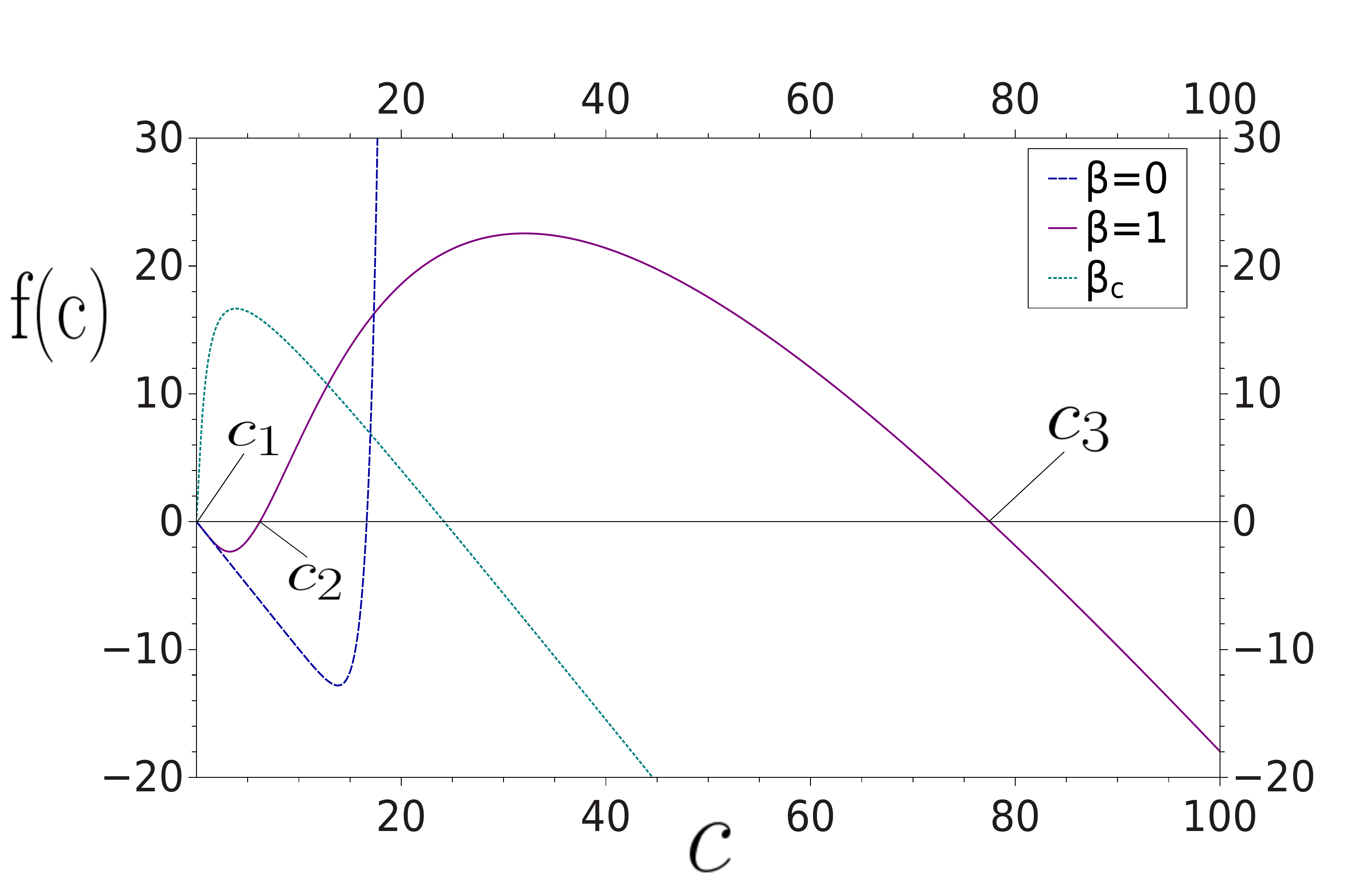}
\caption{ (Color online) Typical forms of the reaction rate f(c) (see Eq.\ref{growthfield}) for different regimes of chemotactic saturation $\beta$. The case $\beta=1$ is plotted with the 
same parameters and with the same exemplary value of $ \Gamma_{\text{global}}(c) $  
as shown in Fig. \ref{deffgrowth}b. In this regime with $0<\beta<\beta_c=\frac{\kappa}{4D\gamma}$ (see Eq. \ref{stabcond}) we typically find three zeros 
$c_{i}$ of f(c). As the global fitness is time dependent, we note that the zeros of f(c) are in general time-depndent: $c_i(t)$.  
For $\beta=0$ there are typically two zeros and for $\beta>\beta_c$ only one zero can be found. For a locally vanishing concentration 
$c({\bf r},t)=0$ the reaction rate $f(c)$ approaches a finite value $f(c)=\frac{d_c}{ \Gamma_{\text{global}}}\exp(-\alpha/\beta)$, which depends on the global fitness.}
\label{formfc}
\end{figure}
%%%%%%%%%%%%%%%%%%%%%%%%%%%%%%%%%%%%%%%%%%%%%%%%%%%%%%%%%%%%%%%%%%%%%%%%%%%%%%%%%5
%%%%%%%%%%%%%%%%%%%%%%%%% FIGURE %%%%%%%%%%%%%%%%%%%%%%%%%%%%%%%%%%%%%%%%%%%%%%%%%%%%%%%%%%%%%%%%
\begin{figure*}
\begin{center}
\includegraphics[width=0.95\textwidth]{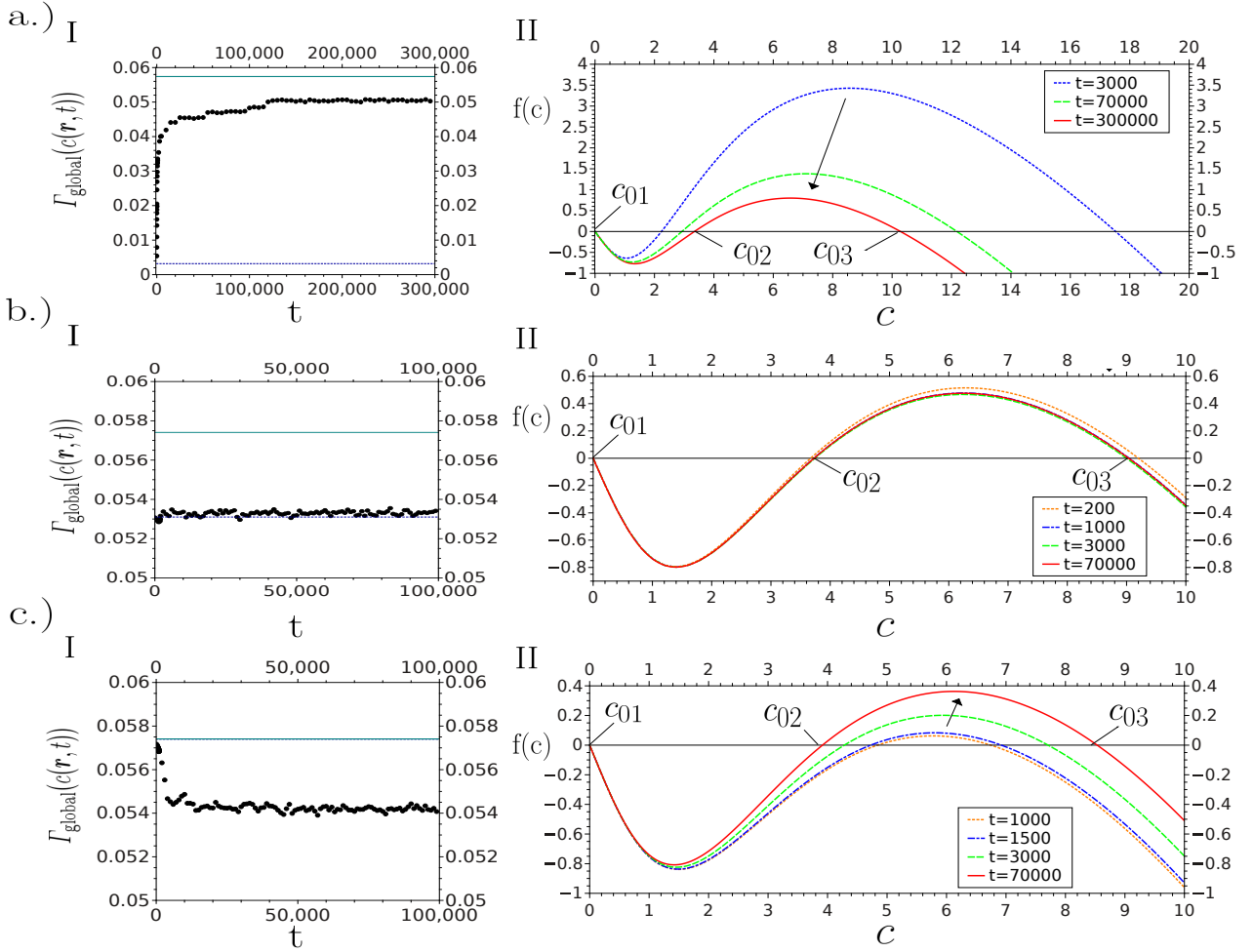}
\caption{ (Color online) Numerical calculation of the global fitness $\Gamma_{\text{global}}(c({\bf r},t))$ (see Eq. \ref{globalgrowthrate}) and reaction rate $f(c({\bf r},t))$ 
(see Eq. \ref{growthfield}) for the long-time simulations shown in Fig. \ref{longtimeplot}. 
(a) Low mean concentration with cluster formation; (b) Intermediate mean concentration with labyrinthine-patterns; (c) High mean concentrations with bubble formation.
Left side (I); Black dots: numerically calculated value
of the global fitness. Blue dashed line: $\Gamma(\overline{c})$. Dark cyan line: $\Gamma(c_+)$. Right side (II); Reaction rate $f(c)$ at the time-s of the snapshots shown in 
Fig. \ref{longtimeplot}. Simulation parameters: $\alpha=\frac{\kappa}{D\gamma}=16$, $\beta=1.5$, $d_c=1$, $q_c=2.5$, $D_c=5$.}
\label{evolutionglob}
\end{center}
\end{figure*}
%%%%%%%%%%%%%%%%%%%%%%%%%%%%%%%%%%%%%%%%%%%%%%%%%%%%%%%%%%%%%%%%%%%%%%%%%%%%%%%%%5

Figure \ref{formfc} illustrates typical forms of $f(c)$ for different regimes of chemotactic saturation. For $\beta=0$ 
we typically have two zeros and for $\beta>\beta_c$ only one zero occurs. In case of finite chemotactic 
saturation with $0<\beta<\beta_c$ we find three zeros of $f(c)$ as long as
\begin{align}
\label{3nullbedingung}
 \Gamma(c_-)<\Gamma_{\text{global}}(c({\bf r},t)) <\Gamma(c_+).
\end{align}
Please note, that $f(c)$ with $\beta=1$ as depicted in Fig. \ref{formfc} corresponds directly to the local and global fitness 
shown in Fig.\ref{deffgrowth}b. The zeros of $f(c)$, corresponding to the dynamical fixed points of the selection equation, are in general time-dependent ($c_{1}(t)$, $c_{2}(t)$, $c_{3}(t)$), because $ \Gamma_{\text{global}}(c({\bf r},t)) $ is time-dependent. 
For $t\to\infty$ the system reaches a stationary inhomogeneous state in which only small fluctuations of the global patterns occur. This corresponds to small fluctuations of 
the global fitness around a stationary value and implies that for $t\to\infty$ the fixed points $c_i(t)$ evolve toward stationary values $c_{01}$, $c_{02}$ and $c_{03}$. During 
the evolution of the system, the values of $c_i(t)$ fulfill the relations 
$$c_1(t) < c_-, \quad c_-<c_2(t)  <c_+, \quad  c_+< c_3(t)$$
with respect to the extrema of the local fitness as long as Eq. \ref{3nullbedingung} holds. 
If we consider a perturbation around $c_i(t)$, linear stability depends on the sign of the derivatives of local and global fitness (linearization of Eq. \ref{reducedfield} around the zeros leads to an expression with a similar structure as Eq. \ref{homogenstab}, but 
without the diffusion term). 
With the same arguments as above, we assume that the global fitness remains constant for small perturbations, so that only the sign of the derivative of the local fitness determines stability around the zeros. This brings us to
\begin{subequations}
\begin{align}
\label{timezeros}
c_1(t)\overset{t\to\infty}{\rightarrow}c_{01} &< c_- : \text{stable} \\
c_-<c_2(t)\overset{t\to\infty}{\rightarrow}c_{02} &< c_+ : \text{unstable} \\
c_3(t)\overset{t\to\infty}{\rightarrow}c_{03} &>c_+  : \text{stable}. 
\end{align}
\end{subequations}
Motivated by the qualitative analogy of the patterns shown in Section \ref{patternformation} to equilibrium liquid-vapor phase transitions, we may use this to distinguish two different regimes: 
\begin{enumerate}
 \item Spinodal decomposition: For $c_-<\overline{c}<c_+$ we are in an unstable regime with respect to the linear stability analysis 
(and obtain a negative effective diffusion coefficient).
\item Metastability: For $\overline{c}<c_-$ or $c_+<\overline{c}$ the homogeneous 
solution is stable with respect to the linear stability analysis, but if 
$\Gamma(c_-)<\Gamma_{\text{global}}(\overline{c}) <\Gamma(c_+)$, we find two stable
zeros $c_1(t)$ and $c_3(t)$ of the reaction rate f(c) and are therefore still in a metastable regime where $c({\bf r},t)$ can evolve from $c_1$ to $c_3$ for supercritical fluctuations.
\end{enumerate}
The above introduced regimes are not meant as exact definitions, but as a semi-quantitative extension of the results of the linear stability analysis; The bubbles shown in 
Fig. \ref{longtimeplot}c emerge for a mean density $\overline{\rho}=2.3$ 
($\overline{c}=5.75$) slightly above the critical density $\rho_{+}=2.28$ ($c_+=5.7$) obtained from the linear stability analysis and show typical behaviour of phase-transition via nucleation. But in general bubbles may also occur for densities $\overline{\rho}<\rho_+$, in the vicinity of $\rho_+$.  
We performed a number of numerical simulations with initial conditions corresponding to finite nuclei (bubbles). Hereby, we observed that these bubbles remain stable in systems with an overall mean density $\overline{\rho}$ clearly above $\rho_+$. This additionally indicates the metastable character. 
The labyrinthine structures in Fig. \ref{longtimeplot}b on the other hand emerge only within the unstable regime, and show typical behaviour of phase separation via spinodal decomposition. The clusters at low densities are formed within the unstable regime but also partly for $\rho<\rho_-$ (see Fig. \ref{stabsimulations}), as the microscopic simulations can easily lead to nonlinear fluctuations at very low mean densities and can lead to a fast growth of localized structures. 

Coming back to the expression in Eq. \ref{growthfield}, we note that we deal with a reaction-diffusion equation with a bistable reaction rate $f(c)$ with two time-dependent stable attractors, $c_1(t)$ and $c_3(t)$, that evolve towards the stable fixed points $c_{01}$ and $c_{03}$ as the stationary inhomogeneous state is reached.  
For the three different initial densities shown in Figure \ref{longtimeplot}, we calculated the time evolution of the global fitness and $f(c)$ numerically. 
Figures \ref{evolutionglob}-I show the time evolution of $ \Gamma(c(t)) $ compared to $\Gamma(c_+)$ and $\Gamma(\overline{c})$, and Figures \ref{evolutionglob}-II show the corresponding reaction rate $f(c)$ at the time-s of the snapshots in Fig. \ref{longtimeplot}. 
Recalling the different global time scales of our model, we note that all time s shown in Fig. \ref{evolutionglob} correspond to times $t>5\tau$ with approximately constant total amount of chemoattractant and only local redistribution of $c({\bf r},t)$. 
In the system with low mean initial concentrations, spots of high concentration (particle clusters) grow and merge 
(see Fig. \ref{longtimeplot}a). 
The spatial redistribution of the concentration $c({\bf r},t)$ that goes along with the merging of the clusters, leads to an increase in the global fitness, which eventually approaches a stationary value below $\Gamma(c_+)$ (see Fig. \ref{evolutionglob}a).

The time-periods of nearly constant global fitness in Fig. \ref{evolutionglob}a for $20.000<t<150.000$
correspond to states where only a few clusters are left; for large systems with only few clusters  very long times may be necessary until two clusters merge and in these time-windows there 
is no significant change in global fitness. 
From the perspective of the analogy of Eq. \ref{reducedfield} to the Eigen-Fisher selection equation (introduced above), these time-windows of nearly stationary global fitness may be 
interpreted as punctuated equilibria \cite{Eldredge1972b} known from evolutionary biology: Evolution leads to selection and the survival of the fittest (selection equation of the Eigen-Fisher 
type finally leads to an inhomogeneous stationary state), and the evolution of the system is characterized by long periods with no significant changes (stationary global fitness) and short, rather 
sudden periods, in which new species evolve (when two clusters merge). Similar characteristics can be observed in evolutionary learning processes of small recurrent networks \cite{Filk2008}. 
The analogy to evolutionary biology has of course its limits, as we are not looking at the formation of new ``species'' (spots) but rather an inverse process of merging of distinct chemotactic spots.
  
At higher densities, corresponding to the labyrinthine structures in Fig. \ref{longtimeplot}b, the global fitness shows only small fluctuations (see Fig.\ref{evolutionglob}b) during the temporal evolution of the pattern (shown in Fig. \ref{longtimeplot}b). Thus, in this regime, the selection dynamics remains in an evolutionary equilibrium.

For even higher densities of particles, close to the critical line $\rho_+$, where bubbles form and merge (see Fig.\ref{longtimeplot}c), we observe that the global fitness 
(slightly) decreases to a stationary value below $c_+$ (see Fig. \ref{evolutionglob}c).
For all three initial densities the global fitness approaches a stationary value around $\Gamma(c_+)$ and the systems evolve into a stationary inhomogeneous state 
of one single cohesive domain of high (low) concentration embedded in a region of low (high) concentration. 

We will now consider growth of a single domain of high concentration (cluster of particles); Recalling the bistable reaction rate $f(c)$, we interpret growth of a cluster in radial 
direction from the center as a front like growth from the higher, stable fixed point $c_3(t)$ to the lower stable, fixed point $c_1(t)$. 
In analogy to  domain growth in non-equilibrium bistable systems (see e.g. \cite{LSGMalcho1985,LSG1992}), we calculate a time dependent critical radius of the cluster. 
We change into polar coordinates, and consider a system with a single cluster at the origin. Assuming a radially symmetric cluster in a 2 dimensional system, we continue to 
investigate the reaction-diffusion-equation (Eq. \ref{growthfield})
that results from the quasi-stationary assumption for the particle density, and rewrite it in polar coordinates  
\begin{align}
\label{startsingle}
\frac{\partial c}{\partial t}= f(c)  + \frac{D_c}{r}\frac{\partial c}{\partial r} + D_c\frac{\partial^2 c}{\partial r^2}.
\end{align}
Here, the reaction rate $f(c)$ is given by Eq. \ref{growthfield}. According to the zeros of the reaction rate (see Eq.\ref{timezeros}), we formulate the boundary conditions:
\begin{subequations}
\begin{align}
 c(0,t)&=c_3(t)>c_+\\
c(\infty,t)&=c_1(t)<c_-.
\end{align}
\end{subequations}
The radius of the saturated domain (single cluster) $R(t)$ can be implicitly defined by
\begin{align}
\label{radiusimplicit}
 c(R(t),t)=c^*
\end{align}
where $c_-<c^*<c_+$ is a fixed concentration in the unstable transition region connecting both of the stable states. Taking the derivative of Eq. \ref{radiusimplicit} with respect to time
\begin{align}
 \frac{\partial c(R(t),t)}{\partial t}+\frac{\partial c(r,t)}{\partial r}\bigg|_{r=R(t)}\frac{d R}{d t}=0
\end{align}
we obtain an expression for the time evolution of the radius of the cluster
\begin{align}
 \frac{dR}{dt}=-\frac{\partial c(r,t)/\partial t}{\partial c(r,t)/\partial r}\bigg|_{r=R(t)}.
\end{align}
If we assume a front-like profile of $c(r,t)$ with a sharp transition region at each time-step, such that $\frac{\partial c}{\partial r}=0$ for $r=0$ and for $r\to\infty$, we can use methods for domain growth in bistable systems 
(see e.g. \cite{LSGMalcho1985,LSG1992}) and obtain
\begin{align}
 \frac{dR}{dt}=D_c\left(\frac{1}{R_k(t)}-\frac{1}{R} \right)
\end{align}
where the time dependent critical radius is given by
\begin{align}
\label{crit_rad}
 R_k(t)=\frac{D_c\int_{A}\left(\frac{\partial c}{\partial r}\right)^2d{\bf r}}{\int_{c_1(t)}^{c_3(t)}f(c)dc}.
\end{align}
Here, time-dependence of $R_k(t)$ is due to the temporal evolution of the global fitness, which results in a time-dependence of $c_i(t)$.
Please note that the critical radius diverges for vanishing denominator ($R_k(t)\to\infty$ for $\int_{c_1(t)}^{c_3(t)}f(c)dc \to 0$), and may change its sign depending on the sign of 
the denominator. We decompose the integral in the denominator according to the zeros of the reaction rate 
\begin{align}
\label{fcintegral}
 \int_{c_1(t)}^{c_3(t)}f(c)dc=\int_{c_1(t)}^{c_2(t)}f(c)dc+\int_{c_2(t)}^{c_3(t)}f(c)dc
\end{align}
in order to clarify, that for the typical form of $f(c)$ we obtain a negative contribution from the first integral and a positive contribution from the second integral on the 
right site of Eq. \ref{fcintegral} (compare to Fig. \ref{formfc} and \ref{evolutionglob}). Both terms are time dependent due to the time dependence of the global fitness; If global fitness increases, $c_3(t)$ decreases and $c_2(t)$ increases
(compare to Fig. \ref{deffgrowth} for different values of $ \Gamma_{\text{global}} $ or see Fig.\ref{evolutionglob}). This 
leads to a decreasing  integral over the reaction rate (Eq. \ref{fcintegral}) and therefore an increasing critical radius. For decreasing 
global fitness on the other hand, $c_3(t)$ increases and $c_2(t)$ decreases, so that the integral over the reaction rate increases and the critical radius decreases.

Numerical calculations show that growth of a single cluster goes along with an increase of $ \Gamma_{\text{global}} $ and therefore the critical radius increases,
until the stationary state is reached. In the stationary state the denominator of the critical 
radius is given by
 \begin{align}
 \label{statfc}
 \int_{c_{01}}^{c_{03}}f(c)dc=\int_{c_{01}}^{c_{02}}f(c)dc+\int_{c_{02}}^{c_{03}}f(c)dc.
\end{align}
These calculations may also be applied to the growth of bubbles instead of clusters; For that we only have to set the origin of polar coordinates in the center of a 
bubble and interpret growth of the bubble as a front-like growth from $c_1$ to $c_3$ (as bubbles grow, $c_3(t)$ shifts towards higher concentrations; See: Fig.\ref{evolutionglob}). 
Thus, we identify a positive $R_k(t)$ in Eq. \ref{crit_rad} with a high-density cluster and a negative $R_k(t)$ with a low-density bubble, whereas a diverging 
critical radius corresponds to straight domain boundaries between high and low density regions.
  
The above considerations were confirmed numerically: We computed the integral over the reaction rate in the different stationary inhomogeneous states - clusters, bubbles and labyrinthine 
structures - that are illustrated in Fig. \ref{longtimeplot}. For the clusters shown in Fig. \ref{longtimeplot}a (as well as for the cluster in Fig. \ref{clustergrowth}) integration over the 
reaction rate (Eq. \ref{statfc}) from 
$c_{01}$ to $c_{03}$ leads to a positive value, and therefore a positive critical radius. For the bubble in Fig. \ref{longtimeplot}c we obtain a negative 
critical radius, and for the pattern with a straight transition region at intermediate densities (see Fig. \ref{longtimeplot}b), the integral over the reaction rate is approximately zero. 
This confirms our expectation of a positive boundary curvature for clusters, a negative one for bubbles and a vanishing curvature (diverging critical radius) for the 
inhomogeneous pattern with a straight transition region.

\section{Single clusters}
\label{singleclusters}
%%%%%%%%%%%%%%%%%%%%%%%%%%%%%%%%% FIGURE %%%%%%%%%%%%%%%%%%%%%%%%%%%%%%%%%%%
\begin{figure*}
\begin{center}
\includegraphics[width=0.99\textwidth]{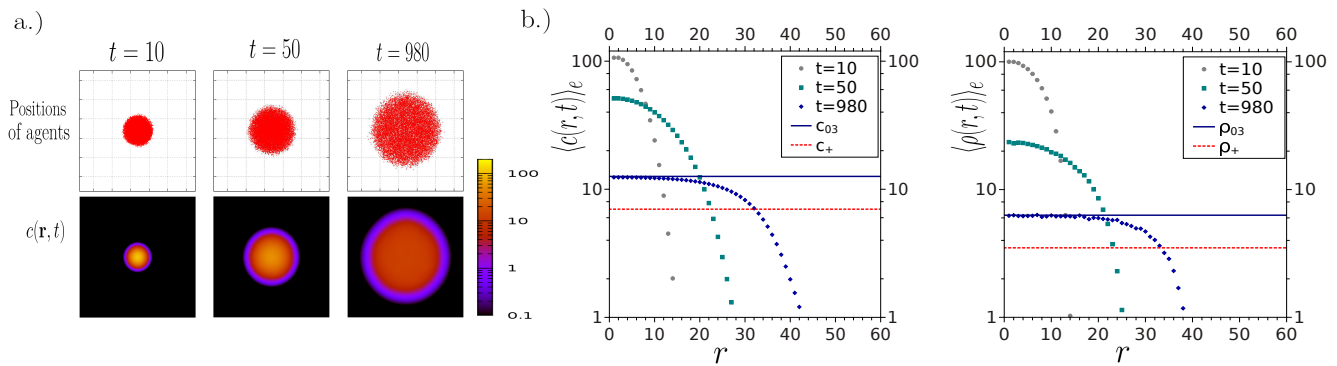}
\caption{ (Color online) Simulation of growth of single clusters. a.) Snapshots at different time s (detail of a $256\times256$ simulation grid) b.) Ensemble averages of the cluster profiles 
$\langle\rho(r,t)\rangle_e$ and $\langle c(r,t)\rangle_e$ at the time-s corresponding to the snapshots (simulation parameters: $\kappa=300,\beta=4,D=2.5,q_c=0.1,d_c=0.05,D_c=0.8$). The value of the stationary fixpoint $c_{03}$ of the reaction rate f(c)
was calculated numerically. Please note, that the density $\rho_{03}$, which corresponds to $c_{03}$ via $\rho_{03}=c_{03}d_c/q_c$ (see Eq. \ref{local}) is also indicated.} 
\label{clustergrowth}
\end{center}
\end{figure*}
%%%%%%%%%%%%%%%%%%%%%%%%%%%%%%%%%%%%%%%%%%%%%%%%%%%%%%%%%%%%%%%%%%%%
\subsection{Domain growth}
If a single cluster is initialized with a density of particles well beyond $\rho_+$, the chemotactic drift is small (due to the saturation for $\beta>0$) compared to the diffusion of the 
particle density and thus the particles spread out -- and the cluster area grows. As $N$ is constant cluster growth is limited and we expect the cluster to finally reach a stationary 
state.

The snapshots in Figure \ref{clustergrowth}a illustrate a typical time evolution of a single cluster initialized with a density of particles well beyond $\rho_+$: As expected the cluster grows until it finally reaches a stationary state.

The snapshots are taken from the simulations in Figure \ref{clustergrowth}b, that show the time 
evolution of the mean profiles $\langle\rho(r,t)\rangle_e$, $\langle c(r,t)\rangle_e$
in radial direction from the center of the cluster. Here $\langle ... \rangle_e$ represents the ensemble-average over $1000$ realizations. 
Initial supersaturation decreases with increasing radius of the cluster until a stationary state is reached. As illustrated in Fig. \ref{clustergrowth}b, numerical calculations of the stationary fixed point $c_{03}$ of the reaction rate f(c) (see Eq. \ref{reducedfield} and 
Eq. \ref{timezeros}) show, that in the stationary state the mean concentration $\langle c(r)\rangle_e$ in the inner region of the cluster approaches the value $c_{03}$. 

In Fig. \ref{clustergrowth}b the density $\rho_{03}$, which corresponds to $c_{03}$ via $\rho_{03}=c_{03}/a$ (see Eq. \ref{local}) is also indicated.  
With increasing distance $r$ from the cluster center, concentration and density of particles in the stationary profiles decrease, and finally approach zero for large $r$. 
In the parameter regime we used, the lower stable fixed point is at very small concentrations ($c_{01}\ll1$), so that the microscopic simulations lead to vanishing values outside the cluster.

\subsection{Stationary characteristics of single clusters}
The reaction rate $f(c)$ has two stationary fixed points, one at $c_{01}<c_-$ and one at $c_{03}>c_+$ and as 
Fig. \ref{clustergrowth} confirms, we can expect a stationary value $\langle c(r)\rangle_e=c_{03}$ in the inner regions of the cluster. In the transition region we can expect $c_{01}<\langle c({\bf r})\rangle_e<c_{03}$ and in the surrounding system 
$\langle c({\bf r})\rangle_e=c_{01}$. 
However, in order to calculate the values of the fixed points, we need to know the stationary value of $ \Gamma_{\text{global}}(c({\bf r})) $, which so far we were only able to calculate numerically.
But based on the discussions of the limiting cases (see Section \ref{limitcases}), we can assume that the stationary profiles
$\langle\rho(r)\rangle_e$ (and $\langle c(r)\rangle_e$) will be distributed around the values $\rho_+$ (and $c_+$)
where the effective diffusion coefficient changes its sign (and we find a maximum of the local fitness). 

%%%%%%%%%%%%%%%%%%%%%%%%%%%%%%%%% FIGURE %%%%%%%%%%%%%%%%%%%%%%%%%%%%%%%%%%%
 \begin{figure*}
\begin{center}
\includegraphics[width=0.8\textwidth]{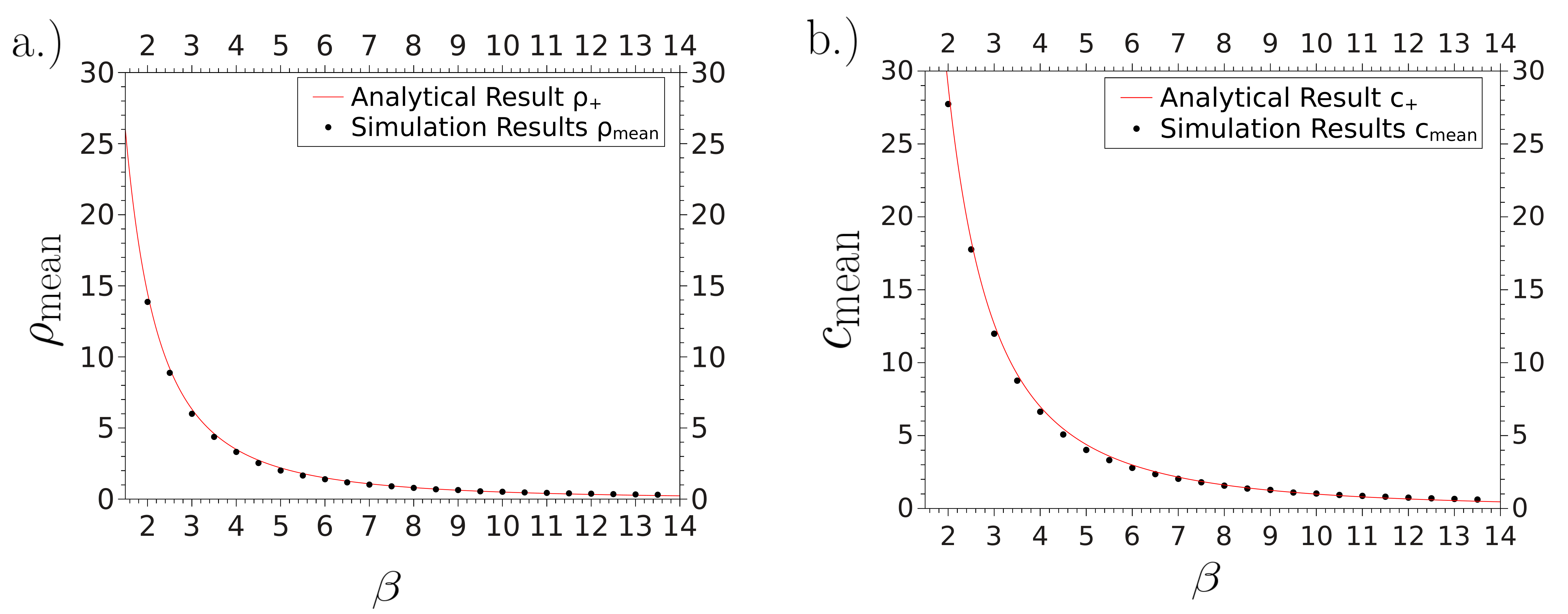}
\caption{ (Color online) Comparison of $\rho_{+}$ and $c_{+}$ to numerical calculations of (a) mean density $\rho_{mean}$ and (b) mean concentration $c_{mean}$ of single clusters in the stationary 
state for different values of the chemotactic saturation parameter $\beta$ (simulation parameters: $q_c/d_c=2,\kappa=300,D=2.5,\gamma=1$. For the 
numerical calculation of the mean values we included the transition region in the outer parts of a cluster)}
\label{cplusrhoplus}
\end{center}
\end{figure*}
%%%%%%%%%%%%%%%%%%%%%%%%%%%%%%%%%%%%%%%%%%%%%%%%%%%%%%%%%%%%%%%%%%%%
%%%%%%%%%%%%%%%%%%%%%%%%%%%%%%%%%%%%%% FIGURE %%%%%%%%%%%%%%%%%%%%%%%%%%%%%%%%%%%%%%%%%%%%%%%%%%%
\begin{figure}
\begin{center}
\includegraphics[width=0.96\columnwidth]{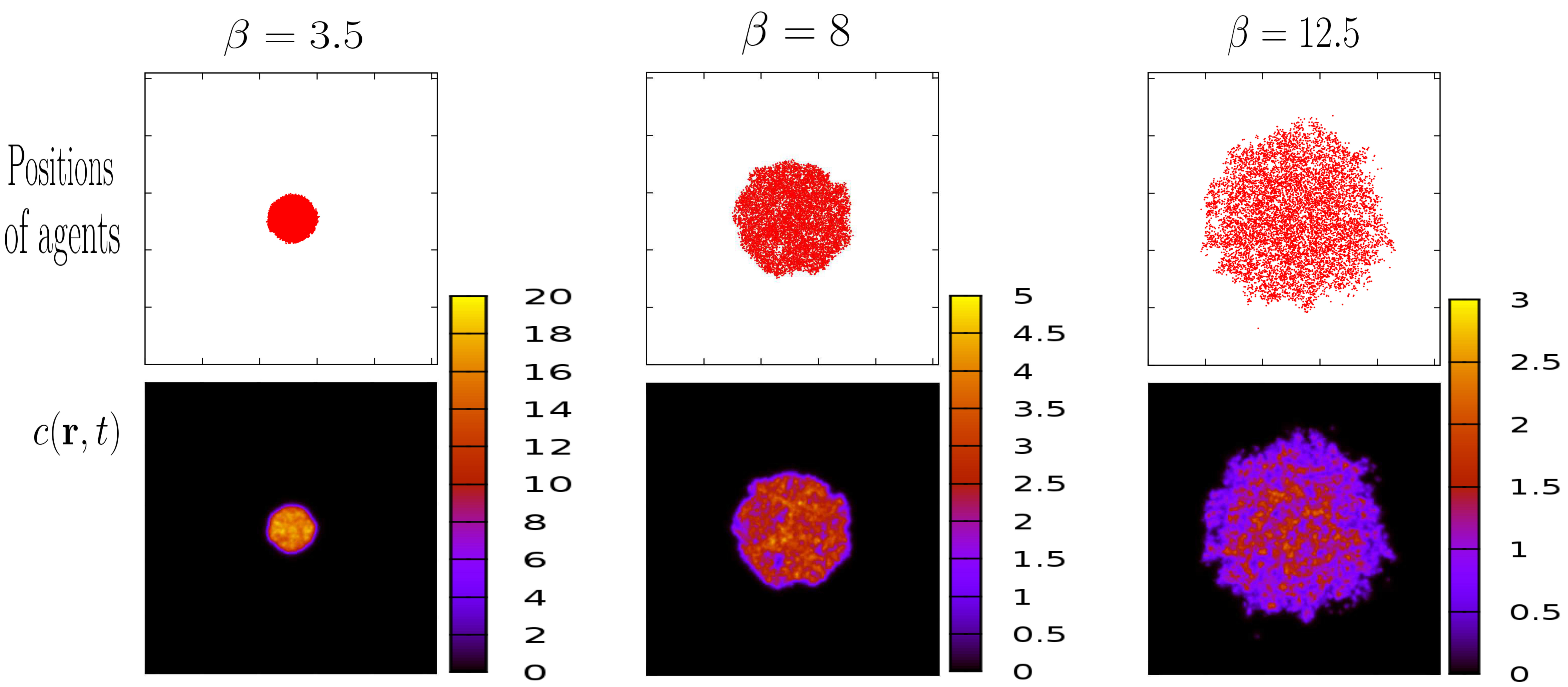}
\caption{ (Color online) Single clusters in the stationary state for different values of the saturation coefficient $\beta$ (snapshots are taken from the simulations shown in Figure 
\ref{cplusrhoplus})} 
\label{cplusrhoplussnapshots}
\end{center}
\end{figure}
%%%%%%%%%%%%%%%%%%%%%%%%%%%%%%%%%%%%%%%%%%%%%%%%%%%%%%%%%%%%%%%%%%%%%%%%%%%%%%%%%%%%%%%%%%%%%%%%

In order to get an approximation of the mean stationary particle density $\rho_{mean}$ (as well as the mean stationary concentration $c_{mean}$) of single clusters, we neglect the actual form of the stationary profiles $\langle\rho(r)\rangle_e$ and $\langle c(r)\rangle_e$ 
and treat a cluster as a disc-like structure of radius $R^0$ with a homogeneous particle density that is given by $\rho_{mean}=\rho_+$ (and $c_{mean}=c_+$) and a sharp transition to a region where the particle density (and concentration of the field) vanishes: 
\begin{subequations}
\begin{align}
\label{meanapprox}
\langle\rho(r)\rangle_e&\approx \rho_{+}\theta(R^0-r)\\
\langle c(r)  \rangle_e&\approx c_{+}\theta(R^0-r)
\end{align}
\end{subequations}
In order to evaluate this approximation, we simulated single clusters in the stationary state for different values of the saturation parameter $\beta$ and numerically calculated $\rho_{mean}$ 
and $c_{mean}$. Hereby we defined the radius of the cluster by including the transition region. As shown in Fig. \ref{cplusrhoplus}, we obtain a very good quantitative agreement between 
analytical approximation and numerical results; The average density (and concentration) of single clusters, 
including the transition region, can be approximated by $\rho_+$ (and $c_+$).
The snapshots in Figure \ref{cplusrhoplussnapshots} illustrate clusters in the stationary state for different values of $\beta$. The cluster size grows with increasing values of $\beta$ and for large $\beta$ single particles start to leave the cluster and its shape and transition region gets irregular.
When we looked closer at the profile of single clusters in the stationary state, we observed dynamic growth and decay of smaller inhomogeneities in the inner regions of the clusters as well as fluctuations in the outer transition region. This underlines the stochastic nature of the stationary clusters. For example, we calculated the mean squared displacement of the center of mass of single clusters of different sizes (numbers of agents) and note that (1) the clusters show normal diffusive behaviour and (2) the diffusion coefficient of large clusters $(N\approx25000)$ approaches the value that one would expect for the center of mass of an 
ensemble of free particles without chemotactic coupling.

\section{Conclusion}
In this work, we have studied the effects of concentration-dependent chemotactic sensitivity
on collective behaviour of autochemotactic Brownian agents. We compared different chemotactic drift functions and showed that the nonlinear Receptor-Law sensitivity leads to two distinct density regimes where the homogeneous solution is stable - a low density regime  ($\langle \rho \rangle <\rho_-$) and a high density regime ($\langle \rho \rangle >\rho_+$). At intermediate densities for  ($\rho_-<\langle \rho \rangle <\rho_+$), and sufficiently strong chemotactic coupling, the homogeneous solution becomes unstable. A variety of spatio-temporal patterns, ranging from bubbles, over labyrinthine-structures to clusters can be observed, and for $t \to \infty$ the system approaches a stable inhomogeneous state with two coexisting phases. In the linearly unstable regime these show characteristics of phase transition via spinodal decomposition, while close to $\rho_{\pm}$ nucleation
processes can be observed. 

The analysis of the limiting cases $\frac{\partial \rho}{\partial t}\approx 0$ and $\frac{\partial c}{\partial c}\approx 0$,  motivates  on the one hand the introduction of an effective density-dependent diffusion coefficient, and on the other hand the introduction of a local (and global) fitness in an equation analogous to a Fisher-Eigen selection equation. Both limits confirm the general result of the linear stability analysis and allow further insights into the dynamical behaviour of the system.

For the Receptor-Law response, the limiting case $\frac{\partial c}{\partial c}\approx 0$ leads to an effective diffusion coefficient that is negative only for a finite range of densities 
$\rho_-<\rho({\bf r},t)<\rho_+$, which correspond to the densities in the transition regions found in the outer parts of large chemotactic clusters. Interestingly, 
agents in the inner regions of such clusters behave as "free" Brownian particles with a positive effective diffusion coefficient, in contrast to the case of constant chemotactic response.    

The limiting case $\frac{\partial \rho}{\partial t}\approx 0$ allows the formulation of an effective bistable, time-dependent reaction-diffusion equation, which was used to obtain results beyond a linear stability analysis. We discussed a metastable regime described by the two stable zeros of the reaction rate, and derived the time-dependent critical radius of a single rotationally symmetric domain, which allows to distinguish stationary patterns, such as high density clusters, low density 
bubbles and phase separated structures with straight domain boundaries. We furthermore showed excellent quantitative predictions on mean stationary characteristics of such structures, as 
for example the averaged density of single clusters.

\bibliography{Bib1105.bib}

\end{document}